# Structures of iron and cobalt bimetallic clusters for optimized chemical vapor deposition growth of single-walled carbon nanotubes

*Qingmei Hu[1], Daniel Hedman[2*], Ya Feng[3*], Wanyu Dai[1], Daisuke Asa[1], Aina Fitó-Parera[4], Yixi Yao[5], Yongjia Zheng[6], Kaoru Hisama[7], Gunjan Auti[1], Hirofumi Daiguji[1], Shohei Chiashi[1], Dmitry Levshov[4], Wim Wenseleers[9], Keigo Otsuka[1], Yan Li[5], Christophe Bichara[8], Sofie Cambré[4], Rong Xiang[6*], Shigeo Maruyama[1,6,10*]*

[1]Department of Mechanical Engineering, The University of Tokyo, Tokyo, 113-8656, Japan

[2]Center for Multidimensional Carbon Materials (CMCM), Institute for Basic Science (IBS), Ulsan, 44919, Republic of Korea

[3]Key Laboratory of Ocean Energy Utilization and Energy Conservation of Ministry of Education, School of Energy and Power Engineering, Dalian University of Technology, Dalian, Liaoning 116024, China

[4]Theory and Spectroscopy of Molecules and Materials, Department of Physics and Department of Chemistry, University of Antwerp, Antwerp 2610, Belgium

[5]Center Beijing National Laboratory for Molecular Science, Key Laboratory for the Physics and Chemistry of Nanodevices, State Key Laboratory of Rare Earth Materials Chemistry and

Applications, College of Chemistry and Molecular Engineering, Peking University, Beijing, 100871, China

[6]State Key Laboratory of Fluid Power and Mechatronic Systems, School of Mechanical Engineering, Zhejiang University, Hangzhou 310027, China

[7]Interdisciplinary Cluster for Cutting Edge Research, Research Initiative for Supra-Materials, Shinshu University, Nagano 380-8553, Japan

[8]CINaM, CNRS - Aix Marseille University, Marseille 13288, France

[9]Nanostructured and Organic Optical and Electronic Materials, Department of Physics, University of Antwerp, Antwerp 2610, Belgium

[10]Institute of Materials Innovation, Institute of Innovation for Future Society, Nagoya University, Furo-cho, Chikusa-ku, Nagoya 464-8603, Japan

* Corresponding authors: Shigeo Maruyama, Daniel Hedman, Ya Feng, Rong Xiang

E-mail address: maruyama@photon.t.u-tokyo.ac.jp; daniel@hedman.science; fengya@dlut.edu.cn; xiangrong@zju.edu.cn.

**ABSTRACT** Bimetallic catalysts play a key role in chirality-selective and efficient growth of single-walled carbon nanotubes (SWCNTs), yet the coupled effects of alloy composition, temperature, and catalyst structure are not fully understood. Here, we investigate iron–cobalt (Fe–Co) alloys as a representative high-performance catalyst system for SWCNT growth in a systematic manner by combining chemical vapor deposition (CVD) experiments with chirality-resolved spectroscopic analysis, as well as molecular dynamics (MD) simulations based on density functional theory–derived machine learning force fields while varying the Fe/Co ratio. Using zeolite-based SWCNTs prepared by alcohol CVD, absorption and photoluminescence spectroscopy, together with two-dimensional excitation–emission fitting was employed to quantify chirality-specific growth efficiency. Two distinct growth regimes were identified. At a relatively low temperature of 600 °C, pure Co exhibits the highest catalytic activity, promoting efficient growth of small-diameter (0.7-0.9 nm) SWCNTs. In contrast, at 850 °C, the $Fe_{0.75}Co_{0.25}$ alloy shows a pronounced enhancement in growth efficiency compared with pure Fe, pure Co, and other Fe–Co compositions, also yielding larger diameter tubes (0.9-1.1nm). Similar growth behavior was observed on $SiO_2$ substrates, enabling detailed transmission electron microscopy analysis of catalyst nanoparticles. Electron microscopy and energy-dispersive X-ray spectroscopy reveal that high SWCNT yields correlate with the formation of small, uniform Fe–Co nanoparticles with Co-enriched surfaces, in excellent agreement with MD simulations. Lastly, MD results are summarized in a composition–diameter phase diagram that rationalizes the experimentally observed growth trends. The exceptional performance of the $Fe_{0.75}Co_{0.25}$ catalyst at high temperature is attributed to the stabilization of small and uniform catalyst clusters, providing mechanistic insight into the synergistic roles of alloy composition and temperature in SWCNT growth.

## INTRODUCTION

Single-walled carbon nanotubes (SWCNTs) have attracted sustained interest owing to their unique one-dimensional structure, exceptional electronic and optical properties, particularly their chirality-dependent electrical conductivity.[1-4] After more than three decades of intensive research, the focus of carbon nanotube studies has gradually shifted from fundamental discovery towards practical applications, such as lithium-ion batteries[5, 6] and nanoelectronic devices.[7-10] Achieving such applications critically relies on the ability to realize both efficient and controlled growth of SWCNTs.[11] Chemical vapor deposition (CVD) represents one of the most mature and widely adopted techniques for SWCNT synthesis.[12-14] In this context, bimetallic catalysts have been extensively employed to achieve high growth efficiency and/or chirality-selective growth. Several recent advances in chirality-controlled SWCNT growth have been realized using binary catalyst systems, such as Co–Mo,[15] $W_6Co_7$,[16] and $Mo_2C$ or WC.[17] More broadly, the roles of binary and ternary alloys, as well as binary metal compounds, have been widely explored for chirality-specific growth.[10] Nevertheless, from a practical perspective, efficient growth remains equally essential, and the fundamental mechanisms governing the performance of bimetallic catalysts—particularly at the level of metal clusters—are still under active debate.

The Co–Mo catalyst system has historically played a central role in the commercial production of CoMoCAT SWCNTs,[18] vertically aligned SWCNT growth,[19] and even chirality-selective synthesis.[15] Despite its widespread use, the origin of its high catalytic efficiency remains unclear. Similarly, Fe–Co catalysts are commonly employed in alcohol CVD growth of SWCNTs.[14, 20, 21] Murakami *et al.*[22] demonstrated that Fe–Co bimetallic catalysts outperform monometallic Fe or Co catalysts at 800 °C, attributing the enhancement to Fe-induced suppression of Co nanoparticle aggregation. However, a general understanding of how alloy composition governs catalytic behavior is still lacking. Subsequent studies by Chiang *et al.*[23] and Xiang *et al.*[24] further revealed that tuning the composition of $Ni_xFe_{1-x}$ and $Co_xMo_{1-x}$ catalysts can effectively modulate the chirality distribution of as-grown SWCNTs. The advantages of alloy catalysts have often been discussed in terms of regulating carbon solubility or suppressing high-temperature sintering, yet direct correlations among alloy composition, catalyst structure, and growth efficiency remain elusive.

In recent years, advances in transmission electron microscopy (TEM) have enabled increasingly detailed structural characterization of catalyst nanoparticles.[4, 25] Cui *et al.*[26] demonstrated that introducing Cu into Co catalysts can effectively regulate nanoparticle size, enabling the direct growth of high-quality SWCNTs with sub-nanometer diameters. More recently, Shiina *et al.*[27] systematically examined a broad range of catalyst compositions, including 41 bimetallic and 18 trimetallic systems, and identified NiSnFe as a unique catalyst capable of producing (6, 5) SWCNTs with an enrichment of approximately 95.8 %. These results underscore the promise of rational alloy design as a pathway toward controlled, and even nearly single-chirality, SWCNT synthesis. Nevertheless, most prior studies have been conducted within limited CVD parameter windows, and a comprehensive understanding of the synergistic interplay among alloy composition, growth temperature, and catalyst structure is in urgent need. Recently, Ru–Co systems have also been extensively investigated by Zhang *et al.*[28] and Everhart *et al.*[29], where efficient growth under specific conditions has been discussed in terms of Mackay-type cluster structures[29] or electronic-structure-derived catalytic activity.[30]

In this work, we focus on Fe–Co alloys, which are among the most widely used and consistently efficient catalyst systems for SWCNT growth. Among various alloy catalysts, Fe–Co bimetallic catalysts have demonstrated superior performance compared with monometallic counterparts over a wide range of CVD conditions.[20] Notably, this composition corresponds to the well-studied wider CVD growth condition with an Fe/Co ratio of 1.[20] In our previous classical molecular dynamics (MD) study, Fe–Co clusters were reported to form a highly symmetric $Fe_{42}Co_{13}$ Mackay structure with an exceptionally high melting point.[31] In this structure, Co atoms occupy the cluster center and the 12 vertices of the outer icosahedron (v-Mackay configuration). Subsequent analysis revealed that the formation of this special structure was strongly influenced by the specific Fe–Co interatomic potential employed, motivating further investigation using DFT-based machine learning force fields.

Here, we establish a comprehensive investigation by integrating CVD growths, atomic-scale catalyst characterizations, and atomistic simulations of catalyst nanoparticles. Using $Fe_xCo_{1-x}$ nanoparticles as catalysts, we systematically examine the influence of alloy composition on SWCNT growth efficiency and diameter distribution across a broad CVD temperature window. Spectroscopic analysis reveals that at a relatively low growth temperature of 600 °C (together with an ethanol partial pressure of 50 Pa, unless otherwise specified), pure Co catalysts yield the highest SWCNT production, with diameters predominantly in the range of 0.7–0.9 nm. Increasing Fe content leads to larger catalyst particle sizes, resulting in broader diameter distributions and a shift towards larger diameters. In contrast, at a higher temperature of 850 °C, the synthesized SWCNTs have a larger diameter of 0.9-1.1nm, and the $Fe_{0.75}Co_{0.25}$ catalyst exhibits a markedly enhanced growth efficiency, significantly outperforming pure Fe, pure Co, and other $Fe_xCo_{1-x}$ compositions.

Furthermore, Raman analysis reveals that SWCNT growth on $SiO_2$ substrates follows trends similar to those observed in zeolite-based systems, enabling detailed TEM characterization of supported catalyst nanoparticles on $SiO_2$ TEM grids. In parallel, we develop a Neuroevolution potential (NEP) machine learning force field based on spin-polarized DFT data for the Fe–Co system and perform extensive cooling MD simulations to identify stable cluster structures. Neither TEM observations nor MD simulations reveal the previously proposed $Fe_{42}Co_{13}$ Mackay structure. Instead, combined TEM and energy-dispersive X-ray spectroscopy (EDS) analyses demonstrate that high SWCNT yields correlate with the formation of small, uniform Fe–Co nanoparticles featuring Co surface segregation, in excellent agreement with MD simulations. Finally, the extensive MD results are summarized in a composition–diameter phase diagram, which rationalizes the experimentally observed growth behavior. In particular, the exceptional performance of the $Fe_{0.75}Co_{0.25}$ catalyst at 850 °C is attributed to the stabilization of especially small and uniform catalyst clusters.

## RESULTS AND DISCUSSION

### Influence of Fe/Co ratio on SWCNT growth – yield and diameter of SWCNTs

To systematically elucidate the influence of the Fe/Co atomic ratio on catalytic behavior during SWCNT growth, we first compared the growth outcomes across a broad operational window of alcohol catalytic chemical vapor deposition (ACCVD). Two representative growth regimes were selected for detailed investigation, as illustrated in **Fig. 1**: (i) a low-temperature, low-pressure regime (600 °C, 50 Pa; milder condition) and (ii) a high-temperature, low-pressure regime (850 °C, 50 Pa; harsher condition). This selection was guided not only by preliminary screening results (**Fig. S1a, S4–S6**), in which absorption spectroscopy was used as a measure for growth yield (**Fig. S1b** and **Fig. S3**) [20] . These results indicated that Fe tends to enhance growth efficiency under harsher conditions whereas Co exhibits higher activity under milder or optimized conditions. Beyond these empirical observations, the selection was also motivated conceptually: harsher environments often amplify subtle differences in catalyst stability, structural reconstruction, and activation pathways, thereby enabling the identification of condition-dependent catalytic specificity that may remain obscured under mild conditions. Such specificity is of particular interest for exploring the origins of chirality-sensitive growth, as gentle growth environments generally offer limited leverage to modulate the energetics required for selective activation. In addition, the extremely low yield obtained at 900 °C, 50 Pa (absorbance < 0.1) rendered further analysis impractical, justifying our focus on the two regimes defined above.

Under the milder condition (600 °C, 50 Pa; **Fig. 1a**), the absorption spectra of SWCNTs synthesized using different $Fe_xCo_{1-x}$ catalysts (except for pure Fe) displayed nearly identical relative peak distributions. For example, the pronounced peak near 983 nm—mainly attributed to the (6, 5) chirality—remained dominant regardless of Fe/Co ratio, as long as Co was present in the catalyst. Monometallic Fe exhibited negligible catalytic activity, while pure Co produced the strongest absorbance, indicating that Co atoms possess intrinsically higher catalytic activity and that Fe incorporation suppresses this activity. Moreover, catalysts containing more than 50 % Fe showed different relative intensities of absorbance peaks compared to catalysts with lower Fe content (**Fig. 1c**), consistent with the chirality-resolved photoluminescence-excitation (PLE) maps (**Fig. S7**). The proportion of longer-wavelength peaks, corresponding to larger-diameter SWCNTs, increased with Fe content.

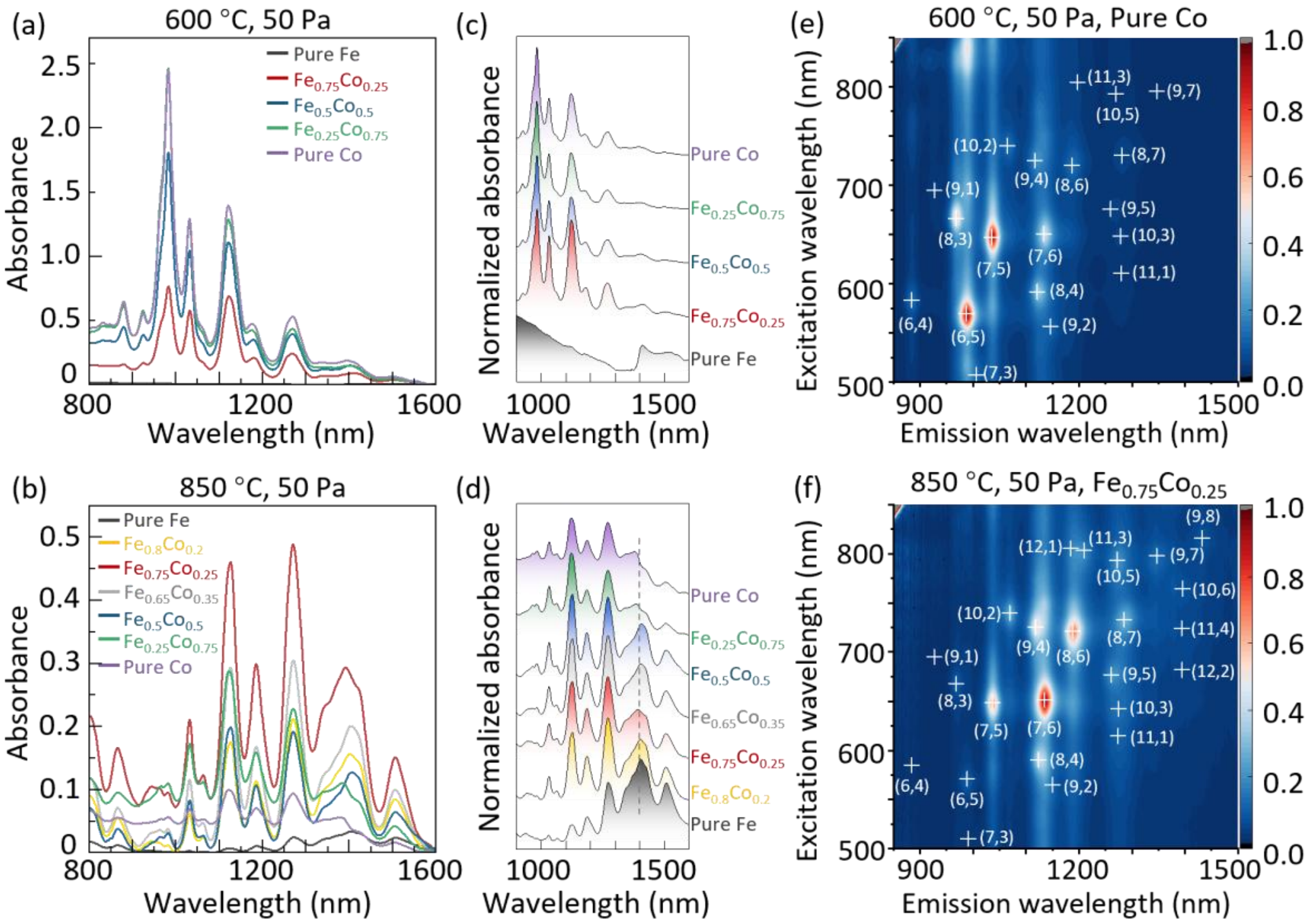

**Figure 1.** Optical characterization. The absorption spectra of dispersed solution samples (following the procedure in **Fig. S2**) grown at (a) 600 °C and (b) 850 °C. Panels (c) and (d) show the corresponding normalized absorption spectra highlighting the consistency in chirality composition. The PLE maps of samples grown with (e) pure Co catalysts, at 600 °C and (f) $Fe_{0.75}Co_{0.25}$ catalysts, at 850 °C, with superimposed chirality identification.

At 850 °C (**Fig. 1b**), in contrast to the 600 °C case, monometallic Fe exhibited measurable catalytic activity, albeit with a modest yield. Among the bimetallic systems, $Fe_{0.75}Co_{0.25}$ produced the highest yield. Compared with the growth behavior of pure Co at both 600 °C and 850 °C, the $Fe_{0.75}Co_{0.25}$ catalyst favored larger-diameter SWCNTs (**Fig. S9**), attributable to both the elevated temperature and increased Fe fraction. As shown in **Fig. 1d**, all catalysts facilitated the growth of large-diameter SWCNTs, and the absorbance peak distribution—except for the band at 1300–1450 nm—remained largely consistent across samples with varying Fe/Co ratios.

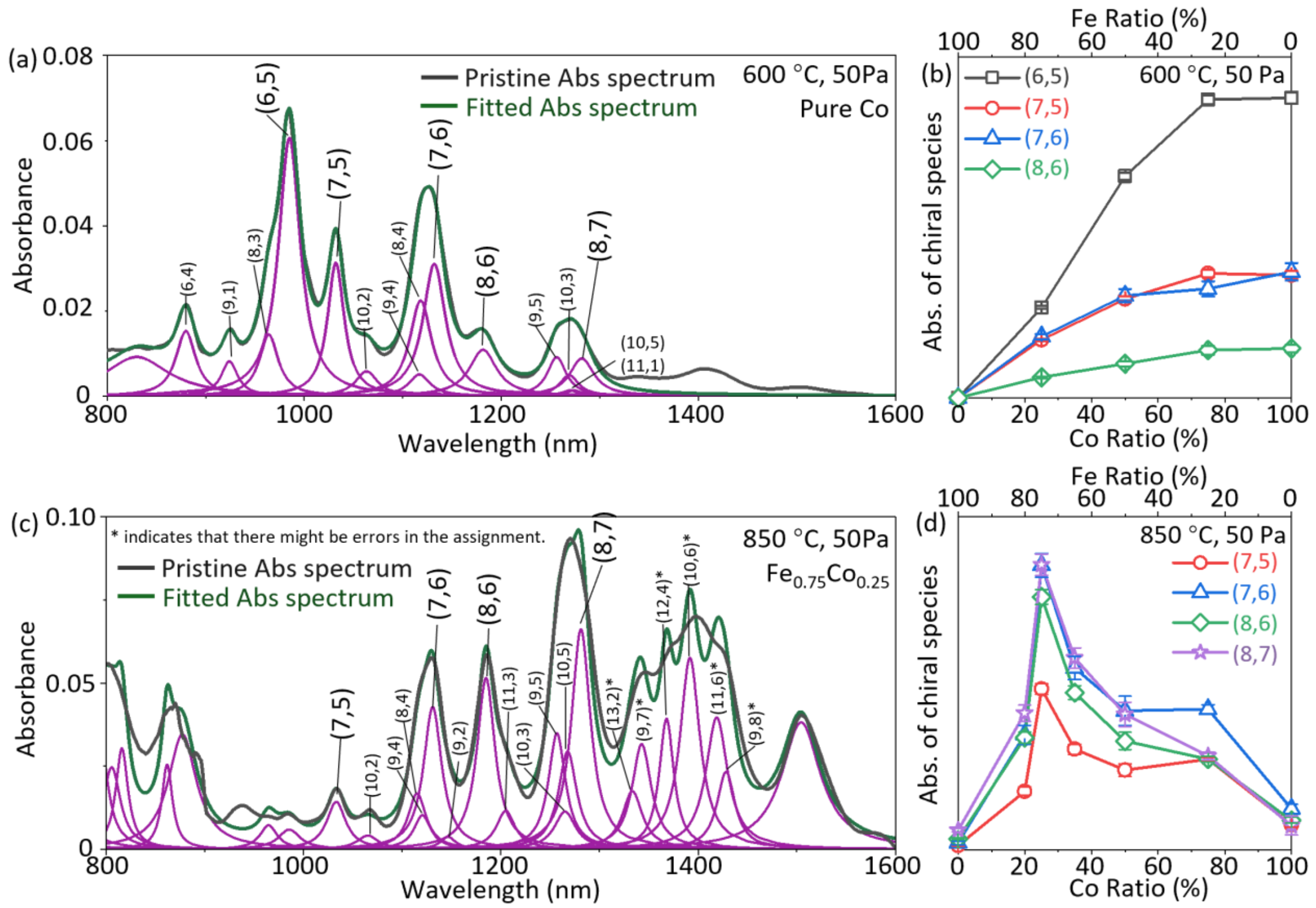

**Figure 2.** Analysis of absorption spectra of the as-grown SWCNT samples that were dispersed in aqueous DOC solutions. (a, c) Fitted absorption spectrum with chiral assignments for the sample prepared using (a) pure Co catalysts grown at 600 °C and (c) $Fe_{0.75}Co_{0.25}$ catalyst grown at 850 °C. All parameters of fitted absorption spectrum are calculated from the fitted PLE mapping, the detailed procedure is shown in the SI. The absorption feature near 1400 nm involves multiple overlapping chiral peaks. Due to limited intensity of these peaks in the current PLE data, the fitted peak positions in this region may carry significant uncertainty. However, the results remain adequate for the purposes of this study, as all data within one series of experiments are fitted in exactly the same manner and only intensities are left to vary. (b, d) Absorbance (the integrated

peak area) dependence of each chiral species on the Co ratio in the catalyst for samples grown at (b) 600 °C and (d) 850 °C.

To gain deeper insight into controlled growth behavior, an advanced correlated PLE and absorption fitting algorithm was employed, as shown in **Fig. 2a**, with a detailed comprehensive description provided in the Data processing part and SI **Section 2**. By fitting the absorption spectra of each $Fe_xCo_{1-x}$ sample individually, the absorbance of different chiral SWCNT species within each sample was quantitatively extracted. **Fig. 2b** and **2d** present representative chiralities with strong absorbance signals under growth temperatures of 600 °C and 850 °C, respectively. At 600 °C, a consistent trend emerged: the absorbance of each chiral species exhibited a positive correlation with the Co content in the catalysts. In contrast, at 850 °C, the sample prepared using the $Fe_{0.75}Co_{0.25}$ catalysts consistently displayed the strongest absorbance peaks, highlighting its superior catalytic efficiency for SWCNT growth under harsher conditions. For a detailed discussion on the advantages of Fe atoms in harsh conditions and Co atoms in mild conditions regarding their impact on growth efficiency, we refer to the discussion of **Fig. S6** in the SI. This growth efficiency observed with the $Fe_{0.75}Co_{0.25}$ catalysts suggested that the Fe/Co ratio possibly plays a critical role in modulating the solubility and precipitation dynamics of carbon atoms or the size of the catalyst nanoparticles, thereby significantly influencing the growth efficiency of SWCNTs.

**Substrate influence on SWCNT growth**

TEM characterizations provide direct insights into the atomic structure of catalyst nanoparticles and as-grown SWCNTs, which is essential for understanding the growth process to enhance the growth efficiency in future works. Leveraging recent advancements in imaging techniques,[32, 33] catalyst particles were directly visualized on thin $SiO_2$ films both before and after

SWCNT growth. A key question is whether the substrate material - $SiO_2$/Si or zeolite in this work - affects the resulting SWCNT products, and specifically whether the atomic structures observed by TEM on $SiO_2$/Si correlate with the SWCNT growth yield evaluated on zeolite substrates by absorption. To compare the structural characteristics of SWCNTs grown on different substrates and to verify the feasibility of using $SiO_2$/Si substrates for in-situ growth and subsequent structural characterization, Raman spectroscopy with four excitation wavelengths (488, 532, 633, and 785 nm) was performed on two sets of samples (#1 synthesized at 800 °C and #2 synthesized at 600 °C) grown on both $SiO_2$/Si and zeolite substrates. As shown in **Fig. 3**, the radial breathing modes (RBM) from samples on both substrates exhibit highly consistent profiles, indicating that the chiral distributions of the SWCNTs are essentially similar, regardless of the substrate type. This result suggests that the substrate exerts a minimal influence on the chirality distribution under the CVD growth conditions employed. Notably, SWCNTs grown on $SiO_2$/Si substrates show slightly broader RBM peaks under certain excitation conditions, particularly in the low-frequency region and in sample #2 synthesized at a relatively low temperature of 600 °C. This peak broadening is likely caused by partial nanotube bundling or interactions on the $SiO_2$/Si surface, leading to inhomogeneous broadening effects that affect peak shapes. However, this difference does not significantly alter the overall chirality distribution, nor does it impede reliable spectral analysis. Meanwhile, as shown in **Fig. S10**, the G and D bands are very similar across the two substrates, with quite good and consistent G/D ratios, indicating high structural quality. These results demonstrate that the use of $SiO_2$/Si does not significantly alter the growth behavior or product quality, validating their applicability as a dual-purpose substrate for both growth and subsequent $SiO_2$/Si-based structural characterizations.[34]

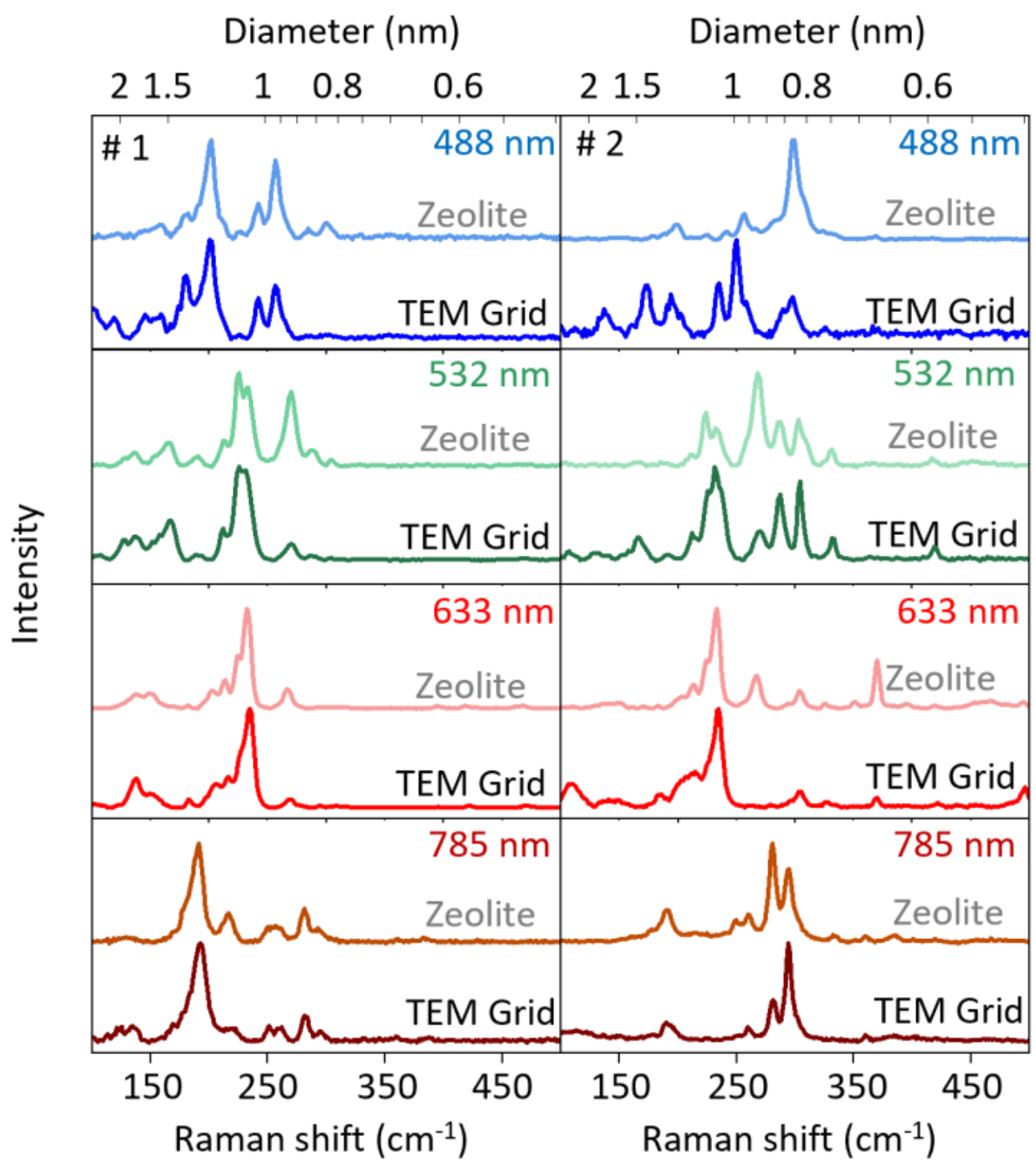

**Figure 3.** Raman spectra (in the RBM range) of two sets of samples (#1 is synthesized with $Fe_{0.5}Co_{0.5}$ at 800 °C, 1.2 kPa and #2 is synthesized with pure Co at 600 °C, 50 Pa, respectively.) grown on zeolite or on a $SiO_2$/Si TEM grid.

## TEM–EDS Characterization and Simulation

Based on the preceding discussions, we prepared varying $Fe_xCo_{1-x}$ catalyst samples (Pure Fe, $Fe_{0.75}Co_{0.25}$, $Fe_{0.5}Co_{0.5}$, and pure Co) under the identical CVD conditions (850 °C) to those used for the zeolite samples from which absorption spectra were obtained. These catalyst samples were deposited on $SiO_2$/Si TEM grids and individually characterized by TEM to investigate their morphology. Subsequently, statistical evaluations were performed to analyze the catalyst NPs. As illustrated in **Fig. 4**, the catalyst NPs were examined both before and after the ACCVD process. This comparative analysis allowed us to assess the structural evolution of the catalyst particles and their influence on SWCNT growth.

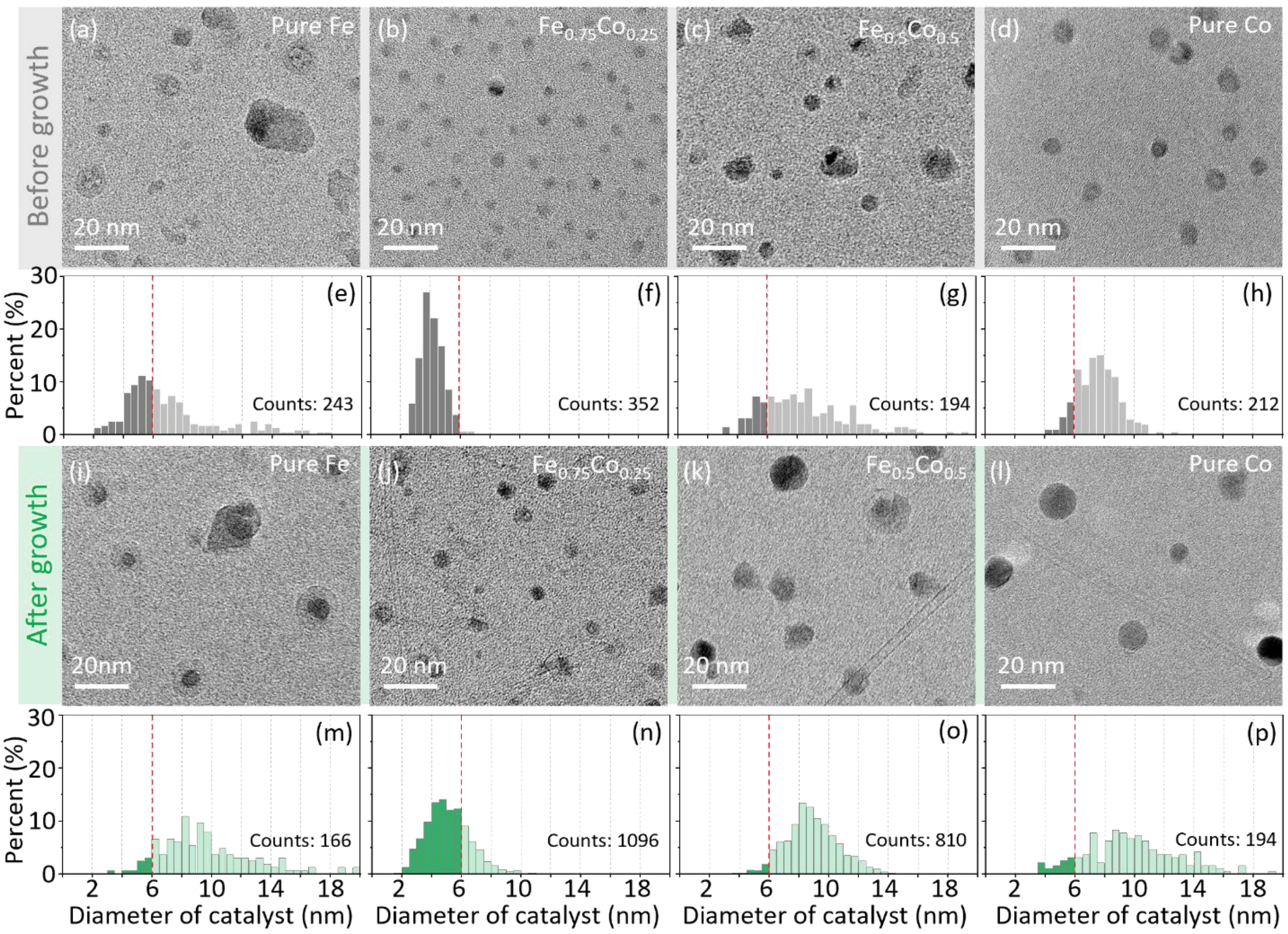

**Figure 4.** TEM characterization and analysis of samples synthesized at 850 °C using different $Fe_xCo_{1-x}$ catalysts. The TEM images of $Fe_xCo_{1-x}$ catalyst particles before (a-d) and after (i-l) SWCNT growth revealing distinct differences in particle sizes. Corresponding histograms of particle size distributions for catalysts before (e-h) and after (m-p) SWCNT growth. Dark-colored bars represent NPs with diameters below 6 nm, which indicates the commonly accepted upper size limit for effective catalysts based on previous studies.[35]

The morphological structure of catalyst particles before ACCVD is a critical determinant of growth behavior. After reduction in Ar/ $H_2$ gas (3 % $H_2$) and prior to growth, the particle size varied significantly depending on the Fe/Co ratio, as illustrated in **Fig. 4a – d**. Specifically, the $Fe_{0.75}Co_{0.25}$ catalyst exhibited a narrow particle distribution and relatively small particle sizes. In contrast, pure Co formed uniform particles of relatively larger sizes. Conversely, the pure Fe catalyst and $Fe_{0.5}Co_{0.5}$ catalyst displayed irregular and heterogeneous particle size distributions.[36] Statistical analysis further corroborated these observations, revealing that the $Fe_{0.75}Co_{0.25}$ catalyst had the smallest average particle size, ranging from approximately 2.5 to 6 nm, as can be seen in **Fig. 4e – h**. Following the ACCVD process, considerable differences in growth efficiency and particle morphology were observed among the catalysts. The $Fe_{0.75}Co_{0.25}$ catalyst demonstrated the highest growth efficiency, which was attributed to its consistently small particle size throughout the process. Traditionally, catalyst particles with diameters exceeding 6 nm tend to lose their catalytic activity for initiating the growth of SWCNTs.[35] Importantly, the particle size of the $Fe_{0.75}Co_{0.25}$ catalyst maintained the smallest even after SWCNT growth, signifying its good carbon transport and superior catalytic performance. For the low-yield samples under these conditions—namely, pure Co and pure Fe—the catalyst particles were found to be encapsulated by an outer layer of deposited carbon after the ACCVD growth process. This behavior is primarily linked to the strong size dependence of carbon solubility in Fe–Co nanoparticles, where larger particles tend to accumulate excess carbon.[37] The resulting carbon coverage provides direct evidence of catalyst deactivation arising from an imbalance between carbon dissolution and precipitation kinetics. These findings collectively highlight the critical influence of catalyst composition—and its impact on particle size—on both growth efficiency and the overall quality of SWCNTs.

Furthermore, we conducted the same statistical analysis on catalyst particles for samples subjected to ACCVD at 600 °C, as presented in **Fig. S11**. In brief, the particle size decreased with increasing Co content prior to ACCVD, leading to higher SWCNT yields after growth, which is consistent with the absorbance trends discussed earlier. A detailed discussion is provided in **SI**.

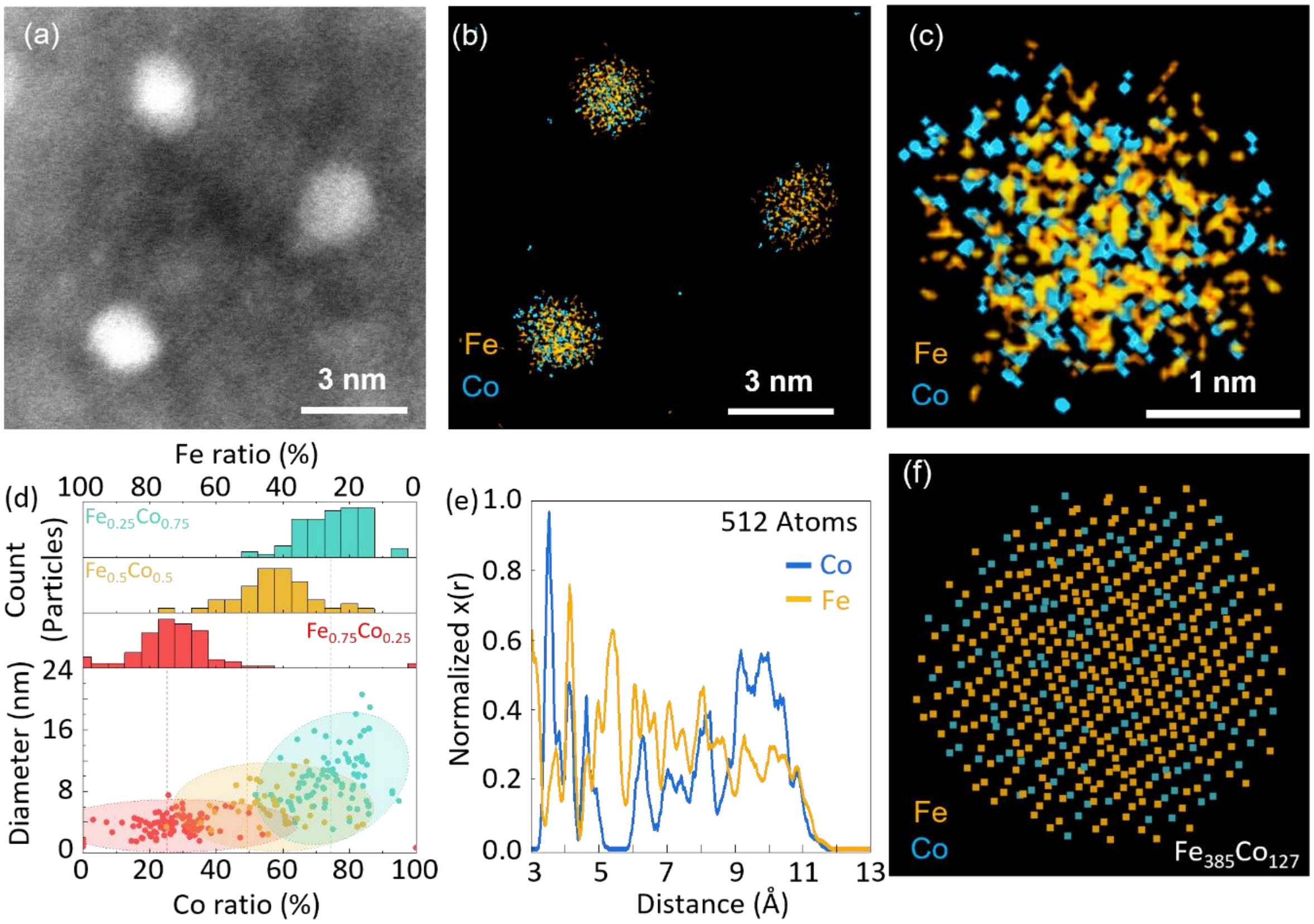

**Figure 5.** Statistical analysis of the Fe–Co ratio in samples grown at 850 °C, obtained by EDS. (a–c) STEM images of $Fe_{0.75}Co_{0.25}$ catalysts after ACCVD growth at 850 °C. Specifically, (a) shows annular dark-field (ADF) images; (b) and (c) presents EDS elemental maps of Fe and Co distributions; (d) Histogram and scatter plot with 95 % confidence region of the experimentally measured Fe–Co ratios in samples grown at 850 °C; (e) Radial concentration profiles of Fe and Co for simulated $Fe_{0.75}Co_{0.25}$ NPs with different total atom numbers, showing size-dependent compositional variations along the radius; (f) Structure obtained from MD simulations with a slow cooling rate, 5 K $ns^{-1}$.

The nominal Co ratio in the catalysts was initially defined by the molar ratio of Fe and Co precursors used during catalyst preparation. To verify the actual $Fe_xCo_{1-x}$ compositions formed under ACCVD conditions, elemental analyses were performed across multiple regions. As shown in **Fig. 5a** and **Fig. S12a**, the ADF-STEM image of the $Fe_{0.75}Co_{0.25}$ sample after ACCVD growth reveals numerous well-defined NPs, and the corresponding elemental maps indicate a strong spatial overlap between Fe and Co signals. This observation confirms that most particles consist of mixed $Fe_xCo_{1-x}$ alloys rather than phase-separated structures. Quantitative analysis of the Fe and Co EDS intensities within individual particles yields calculated Co-ratio maps, shown in **Fig. S12c** (low magnification) and **Fig. S12f** (high magnification). In these maps, yellow corresponds to Fe-rich regions and blue to Co-rich regions.

The majority of particle cores exhibit a continuous color transition from deep yellow to deep blue, indicating well-alloyed $Fe_xCo_{1-x}$ NPs. In contrast, the particle surfaces are predominantly blue, suggesting a Co-enriched surface region. Statistical analysis of particles synthesized from precursors with nominal compositions of $Fe_{0.75}Co_{0.25}$, $Fe_{0.5}Co_{0.5}$, and $Fe_{0.25}Co_{0.75}$ reveals that the $Fe_{0.75}Co_{0.25}$ sample exhibits the narrowest diameter distribution. As summarized in **Fig. 5d**, both the histogram and scatter plot with 95 % confidence region reveal that the experimentally measured Fe/Co ratios closely match the precursor stoichiometry. For example, the $Fe_{0.75}Co_{0.25}$ sample exhibits an average Co fraction of approximately 25 %, confirming that the ACCVD process preserves the intended chemical stoichiometry without significant elemental loss or compositional drift. To elucidate the structure of these NPs, MD simulations were performed on an $Fe_{407}Co_{154}$ NP (corresponding to ~$Fe_{0.75}Co_{0.25}$) under slow cooling conditions (5 K $ns^{-1}$). As shown in **Fig. 6d,** the NP evolves into a BCC Wulff-shaped structure. The radial Fe, Co concentration profile as measured from the center of the NP was obtained after cooling, shown in **Fig. 5e** and **5f.** Here, a pronounced tendency for Co surface segregation is observed, in qualitative agreement with the experimental measurements (**Fig. S12c** and **S12f**). These findings collectively demonstrate that the alloy NPs adopt a mixed Fe-Co BCC structure with an overall Fe-rich composition and a tendency toward Co surface enrichment.

To further elucidate the thermodynamic origins of the composition-dependent catalytic behavior, MD simulations were performed to evaluate the structure and energetics of $Fe_xCo_{1-x}$ NPs with various sizes and compositions (**Fig. 6**). As shown in **Fig. 6a**, the solidification temperature exhibits a size-dependent trend and gradually converges once the NP contains several hundred atoms, indicating that larger NPs solidify through comparable thermodynamic pathways irrespective of composition. For small NPs, the solidification point increases almost linearly with Co content, suggesting enhanced structural rigidity upon Co incorporation. The excess-energy profiles in **Fig. 6b** reveal the energetic competition underlying these trends. Small NPs consistently favor Fe–Co alloying, whereas larger NPs display a pronounced excess-energy maximum near 20–35 % Co—a signature of metastable configurations arising from the competition between Fe-favored BCC and Co-favored close-packed motifs. This competition becomes even more evident when examining the effective surface energy in **Fig. 6c**. As the NP size increases, the effective surface energy develops a clear convex peak in the same 20–35 % Co range, indicating increased compositional sensitivity of the surface and enhanced atomic mobility at elevated temperatures. A possible stronger interaction between surface Co atoms and the substrate are expected to suppress coalescence of NPs and help maintain small catalyst sizes during ACCVD growth.

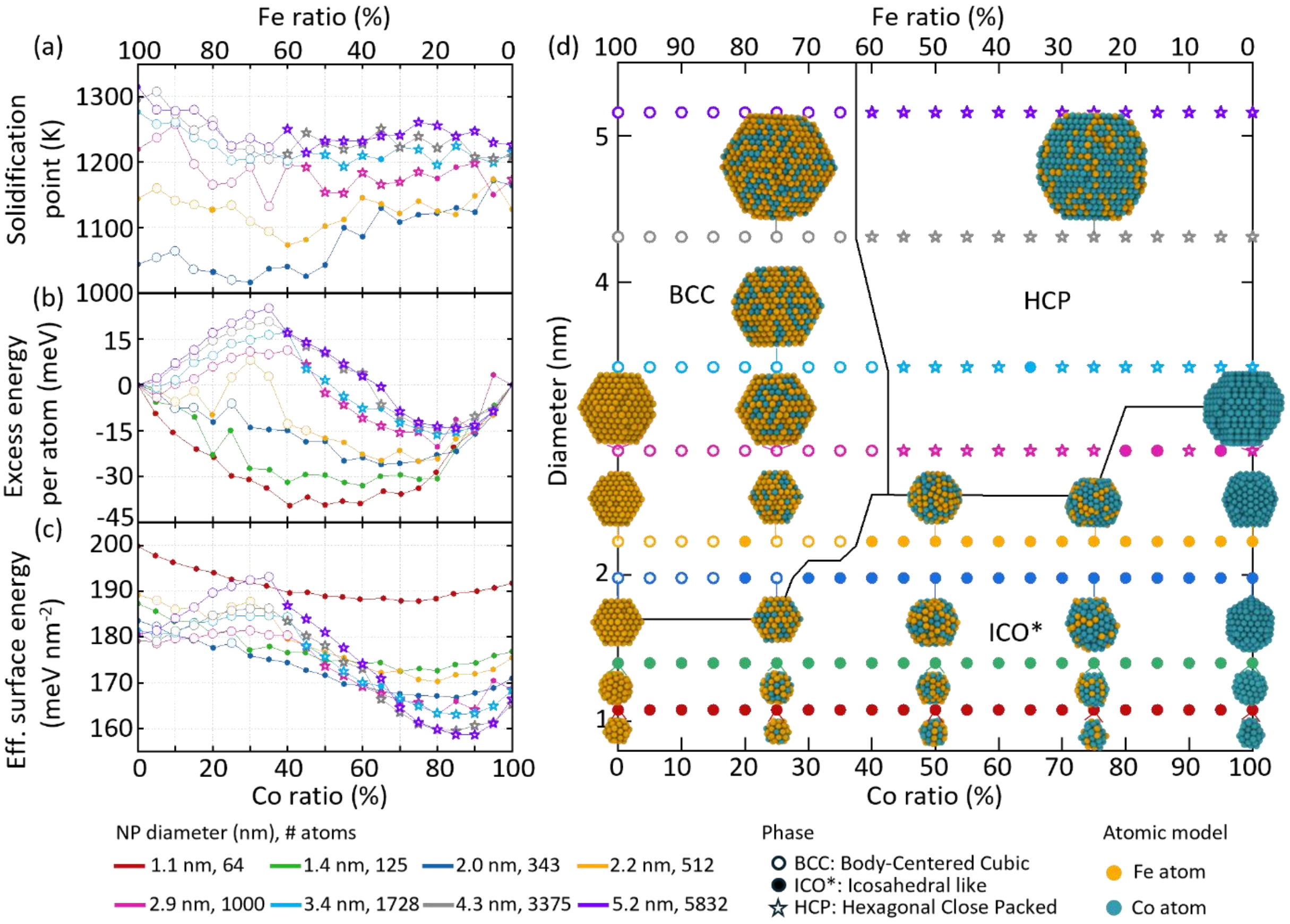

**Figure 6.** Machine learning force field driven MD simulations of slow cooling at 5 K $ns^{-1}$ for different Fe-Co NP sizes. (a) Solidification point, (b) Excess energy, (c) Effective surface energy, and (d) Structural map showing the dominant body centered cubic (BCC), hexagonal close-packed (HCP) and icosahedral like (ICO*) structures across size and composition.

The structural origin of these energetic anomalies is directly verified by the phase map shown in **Fig. 6d**. Small clusters predominantly adopt icosahedral like (ICO*) geometries due to surface-energy minimization, whereas larger Fe-rich and Co-rich clusters evolve toward bulk-like BCC and HCP structures, respectively. Notably, in the intermediate 20–35 % Co composition range—where both excess energy and surface energy exhibit pronounced maxima—the competition among ICO*, BCC, and HCP motifs becomes particularly intense. The elevated surface energy in this regime may also significantly enhance the interaction between NPs and the substrate. Isolated clusters with compositions in this range tend to compensate for unstable surface atoms by forming stronger interfacial bonding with the substrate, thereby achieving improved size stability under growth conditions. In addition, the configurational fluctuations arising from the competitive coexistence of ICO*/BCC/HCP structures may further contribute indirectly to stabilizing the clusters on the substrate. Together, these factors provide a coherent explanation for the experimentally observed small particle sizes, high growth efficiency, and narrow diameter distribution in the $Fe_{0.75}Co_{0.25}$ catalyst system. These structural features are visually demonstrated by the representative atomic models shown in **Fig. 6d** and **Fig. S14**. The models illustrate the evolution of Fe–Co clusters of various sizes (from 64 to 5832 atoms) and compositions, clearly revealing the transition from ICO*-dominated geometries in small clusters to more BCC/HCP-like motifs in larger clusters.

## CONCLUSION

This study systematically demonstrates the synergy between $Fe_xCo_{1-x}$ catalyst composition and ACCVD growth conditions in governing SWCNT growth. Experimental and MD simulation results demonstrate a distinct diameter-dependent growth trend, where smaller NPs favor narrower SWCNTs, and the thermal stability of the NPs plays a dominant role in the final diameter distribution and yield. At lower temperatures (*e.g.*, 600 °C), Co NPs remain well-dispersed and highly active, leading primarily to small-diameter SWCNTs (0.7–0.9 nm). As temperature rises, Co particles coalesce, likely disrupts the carbon dissolution–precipitation equilibrium due to the strong size dependence of carbon solubility and lead to catalyst deactivation. The introduction of Fe effectively suppresses particle growth, enhances high-temperature stability, and promotes the formation of larger-diameter SWCNTs (0.9–1.1 nm) under elevated temperatures. Notably, the optimal Fe content is intrinsically temperature-dependent, with higher growth temperatures requiring a greater proportion of Fe to stabilize the alloy NPs. MD simulations provide microstructural insights into this stabilization mechanism. Within a Co composition range of ~20–35 % (*e.g.*, $Fe_{0.75}Co_{0.25}$), the alloy NPs exhibit significant structural competition (ICO*/BCC/HCP coexistence) and highly dynamic surface reconstruction, helping rationalize the enhanced stability. In summary, this work highlights the importance of precise control over both $Fe_xCo_{1-x}$ catalyst composition and growth temperature, providing clear design principles for scalable and controlled SWCNT synthesis.

## METHODS

### Synthesis of SWCNTs on zeolite

Catalytic Fe and Co were supported on USY-zeolite (HSZ-390HUA, Tosoh) via impregnation with a homogeneous mixture of iron(II) acetate [$(CH_3COO)_2Fe$], cobalt(II) acetate tetrahydrate [$(CH_3COO)_2Co\cdot 4H_2O$] and ethanol, following a procedure analogous to that reported in previous studies.(The total amount of Fe and Co is 5 wt%)[14] Zeolite coated with catalytic Fe and Co NPs at varying molar ratios was evenly distributed on a quartz boat, which was then placed inside a quartz tube (27 mm in diameter) within an electric furnace, as shown in **Fig. S2a**. $Fe_xCo_{1-x}$, the ratio is the molar ratio of Co atoms to Fe atoms in the mixture of $(CH_3COO)_2Co\cdot 4H_2O$ and $(CH_3COO)_2Fe$, the total molar amount of metal atoms is fixed. During the heating process to the target temperature and the subsequent reduction reaction, 300 sccm of $Ar/H_2$ gas (3 % $H_2$) was continuously introduced into the quartz tube, maintaining an internal pressure of 40 kPa. Following the reduction process, the $Ar/H_2$ gas flow was stopped, and the quartz tube was evacuated using a rotary pump. 310 sccm mixture gas of Ar and ethanol vapor, with Ar acting as the carrier gas, was then introduced to keep the target partial pressure of ethanol (typically 50 Pa). Upon completing the CVD reaction, the system was cooled to room temperature under a flow of either Ar or $Ar/H_2$ gas. The growth temperature ranged from 500 °C to 900 °C, and the growth pressure (ethanol partial pressure unless otherwise specified) varied between 10 Pa and 10 kPa. After cooling, the SWCNT sample grown on zeolite in the quartz boat was collected for subsequent characterizations. For detailed information on CVD conditions, please refer to **Fig. S2a** and the previous reports.[14, 20, 33, 38, 39]

**Synthesis of SWCNTs on TEM grid**

Fe and Co catalysts were directly deposited onto a $SiO_2$/Si TEM grid (20 nm $SiO_2$ on the top) using magnetron sputtering (Advance Riko). The grid was then annealed in air at 400 °C for 5 min to fix the catalyst film. After annealing, the grid was transferred into a CVD chamber for SWCNT synthesis, as shown in **Fig. S2b**. After CVD growth, the grid, containing the catalysts only or the catalysts with as-grown SWCNTs, was used directly for TEM characterizations, enabling observation of the original morphology of the catalysts and SWCNTs. TEM characterizations were performed using a JEOL 2010F microscope for general TEM imaging, operated at an acceleration voltage of 200 kV. High-angle annular dark-field scanning TEM (HAADF-STEM) images were captured using the JEM-ARM200F microscope equipped with a cold field-emission gun, employing a probe size typically smaller than 0.1 nm, operated at an acceleration voltage of 80 kV. For additional details on direct observations using a $SiO_2$/Si TEM grid, please refer to the previous reports.[27,35,36]

**Characterizations**

Samples were analyzed using a range of techniques. The solid samples were examined via scanning electron microscopy (SEM, Hitachi S-4800) at an accelerating voltage of 1 kV and a Renishaw inVia Raman system, employing laser excitation wavelengths of 488, 532, 633, and 785 nm. Additionally, dispersed SWCNT solution samples were prepared for absorption and photoluminescence-excitation (PLE) measurements. To prepare these solution samples, 8 mg of as-grown SWCNTs on zeolite powder were dispersed in 5 mL of a DOC/$D_2O$ solution (0.5 wt% sodium deoxycholate (DOC) in deuterium oxide ($D_2O$)) [40] through tip-sonication (power: 400 W; amplitude: 50 %; time: 15 min) and centrifugation (relative centrifugal force (RCF): 11,180 × *g*; time: 3 h), as illustrated in **Fig. S2c**. The UV-vis-NIR absorption and PLE spectra of the SWCNT solutions were recorded using a UV-3150 spectrophotometer (Shimadzu Co., Ltd.) and a HORIBA Jobin Yvon Fluorolog iHR320 equipped with a liquid-nitrogen-cooled InGaAs detector, respectively. To minimize reabsorption effects, the absorbance of the suspensions used for PLE measurements was kept below 0.1 (using a 1 cm pathlength cuvette).

**Data processing**

**Background subtraction of absorption spectra**

The collected raw SWCNT samples inevitably contain some impurities, such as amorphous carbon and fullerenes. These impurities can interfere with the subsequent optical characterizations. Besides, scattering effects also contribute to a background in absorption spectrum given the SWCNTs have a length in the range of the wavelength of the excitation light. As a result, the measured absorbance includes a strong background, indicated by the grey area in **Fig. S1b**. Therefore, prior to comparing the absorption spectra of different samples, it is essential to perform a proper background correction to eliminate the influence from the scatterings and the impurities. In this work, we adopted a background subtraction strategy reported by Tian *et al*,[41, 42] employing a combined Fano and Lorentzian function model to accurately correct the background in the absorption spectra, thereby ensuring the reliability and comparability of the data. The post-processed data are shown in **Fig. 1c** and **1d**, and a detailed description of the fitting process can be found in SI **Section 2**.

**Determination of chirality distribution**

Quantifying the relative abundance of each chiral species in SWCNTs synthesized with different $Fe_xCo_{1-x}$ catalysts is crucial to elucidate the influence of bimetallic composition on controlled growth. Two-dimensional photoluminescence excitation spectroscope (2D PLE) is a powerful technique for characterizing semiconducting SWCNTs. Since the optical emission and excitation transitions of SWCNTs are strongly correlated with their chirality, the peak positions in PLE spectra enable precise identification of specific chiral species. However, PLE spectra alone cannot directly provide the relative abundance of each chirality species due to variations in emission efficiencies, which depend on both chirality and diameter,[43] as well as environmental interactions.[44] Still, comparing PLE spectra measured in the same manner can provide relative changes in chiral species abundances between samples. Likewise, absorption spectra cannot provide absolute quantification of chiral species abundance due to diameter- and environment-dependent absorption cross-sections, however they remain a direct indicator of relative abundance for each (n, m) species. Moreover, because overlapping peaks in the absorption spectra complicate the analysis, the data can only provide an approximate estimate of the overall chirality distribution. Since neither PLE nor absorption spectra alone can accurately estimate chiral species abundance, we combined both methods to achieve a more reliable quantification. Firstly, we systematically fitted the 2D PLE spectra using the method developed by Cambré *et.al*,[45] extracting key spectral parameters such as peak positions and full width at half maximum (FWHM) for each chirality. These parameters were then used to fit the absorption spectra, enabling reliable quantification of the relative abundance of each chiral species (**Fig. 2a and 2c**). Further methodological details are available in the **SI Section 2**.

**Molecular dynamics simulation**

Molecular dynamics (MD) simulations were carried out using the Graphics Processing Units Molecular Dynamics (GPUMD) software[46] version 4.2. To determine the structure and solidification point of each nanoparticle size, slow cooling simulations were performed. Each nanoparticle was cooled from its initial temperature down to 300 K at a rate of 5 K $ns^{-1}$. The initial temperatures were chosen based on nanoparticle size: 1200 K for $M_{64}$, 1500 K for $M_{125}$ to $M_{1728}$ and 2000 K for $M_{3375}$ to $M_{5832}$. After cooling, each nanoparticle was relaxed to obtain its equilibrium structure and free energy.

From the free energy of the relaxed nanoparticles their excess energy, $E_{\mathrm{ex}}$, was calculated as $E_{\mathrm{ex}} = \frac{E_{\mathrm{FeCo}} - N_{\mathrm{Fe}}\, E_{\mathrm{Fe}} - N_{\mathrm{Co}}\, E_{\mathrm{Co}}}{N_{\mathrm{FeCo}}}$. Here $E_{\mathrm{FeCo}}$ is the total energy of the nanoparticle, $N_{\mathrm{FeCo}}$ is the total number of atoms in the nanoparticle, $N_{\mathrm{Fe}}$ and $N_{\mathrm{Co}}$ are the number of Fe and Co atoms in the nanoparticle, respectively and $E_{\mathrm{Fe}}$, $E_{\mathrm{Co}}$ is the energy per atom of pure Fe, Co nanoparticles of the same size. The effective surface energy, $\gamma_{\mathrm{eff}}$, was calculated in a similar way $\gamma_{\mathrm{eff}} = \frac{E_{\mathrm{FeCo}} - N_{\mathrm{Fe}}\, E^{*}_{\mathrm{Fe}} - N_{\mathrm{Co}}\, E^{*}_{\mathrm{Co}}}{4\pi \langle 0.5D \rangle^{2}}$. Here $E^{*}_{\mathrm{Fe}} = -6.090$ eV and $E^{*}_{\mathrm{Co}} = -4.863$ eV is the energy per atom for bulk BCC Fe and HCP Co obtained using the MLFF. $\langle 0.5D \rangle$ is the mean radius for all nanoparticles of the same size (number of atoms).

All MD simulations used a timestep of 2.0 fs and were conducted in the NVT ensemble, with temperature controlled by a Bussi-Donadio-Parrinello thermostat[47] using a coupling constant of 100. Every 100 timesteps, both linear and angular momentum were removed to prevent drift and rotation. A Fe-Co-Si-O neuroevolution potential (NEP)[48-51], still under development, was employed to model interatomic interactions. The NEP is trained on a dataset of diverse structures obtained from active learning and labeled with energy, forces and virials from spin-polarized DFT calculations using the Vienna Ab initio Simulation Package (VASP). Although the detailed description of this potential will be published separately, its reliability was assessed in the present work by using an ensemble of NEPs to estimate the model deviation (the max deviation in the predicted atomic forces for each timestep). For all melting simulations, the mean model deviation remained below 200 meV $Å^{-1}$, which we consider is sufficiently accurate for the current study. All atomic structures obtained from the molecular dynamics simulations were identified and visualized using the OVITO software.[52]

## AUTHOR INFORMATION

### Corresponding Authors

**Shigeo Maruyama** − State Key Laboratory of Fluid Power and Mechatronic Systems, School of Mechanical Engineering, Zhejiang University, Hangzhou 310027, China; Department of Mechanical Engineering, The University of Tokyo, Tokyo 113-8656, Japan; Institute of Materials Innovation, Institute of Innovation for Future Society, Nagoya University, Furo-cho, Chikusa-ku, Nagoya 464-8603, Japan; orcid.org/0000-0003-3694-3070; Email: maruyama@photon.t.u-tokyo.ac.jp

**Daniel Hedman** − Center for Multidimensional Carbon Materials (CMCM), Institute for Basic Science (IBS) (Republic of Korea), Ulsan, 44919, Republic of Korea; orcid.org/0000-0003-1542-6170; orcid.org/0000-0003-1542-6170; Email: daniel@hedman.science

**Ya Feng** − Key Laboratory of Ocean Energy Utilization and Energy Conservation of Ministry of Education, School of Energy and Power Engineering, Dalian University of Technology, Dalian, Liaoning 116024, China; orcid.org/0000-0003-3049-9933; Email: fengya@dlut.edu.cn

**Rong Xiang** − Department of Mechanical Engineering, The University of Tokyo, Tokyo 113-8656, Japan; State Key Laboratory of Fluid Power and Mechatronic Systems, School of Mechanical Engineering, Zhejiang University, Hangzhou 310027, People’s Republic of China; orcid.org/0000-0002-4775-4948; Email: xiangrong@zju.edu.cn

**Authors**

**Qingmei Hu** − Department of Mechanical Engineering, The University of Tokyo, Tokyo 113-8656, Japan; orcid.org/0000-0002-3292-0130

**Wanyu Dai** − Department of Mechanical Engineering, The University of Tokyo, Tokyo 113-8656, Japan; orcid.org/ 0009-0002-9498-1382

**Daisuke Asa** − Department of Mechanical Engineering, The University of Tokyo, Tokyo 113-8656, Japan

**Aina Fitó-Parera** − Theory and Spectroscopy of Molecules and Materials, Department of Physics, University of Antwerp, Antwerp 2610, Belgium; orcid.org/0000-0003-3914-6640

**Yixi Yao** − Center Beijing National Laboratory for Molecular Science, Key Laboratory for the Physics and Chemistry of Nanodevices, State Key Laboratory of Rare Earth Materials Chemistry and Applications, College of Chemistry and Molecular Engineering, Peking University, Beijing, 100871, China; orcid.org/0000-0001-5406-2037

**Yongjia Zheng** − State Key Laboratory of Fluid Power and Mechatronic Systems, School of Mechanical Engineering, Zhejiang University, Hangzhou 310027, People’s Republic of China; orcid.org/0000-0001-5836-6978

**Kaoru Hisama** − Interdisciplinary Cluster for Cutting Edge Research, Research Initiative for Supra-Materials, Shinshu University, Nagano 380-8553, Japan; orcid.org/0000-0002-0508-6427

**Gunjan Auti** − Department of Mechanical Engineering, The University of Tokyo, Tokyo 113-8656, Japan; orcid.org/0000-0003-1474-1670

**Hirofumi Daiguji** − Department of Mechanical Engineering, The University of Tokyo, Tokyo 113-8656, Japan; orcid.org/0000-0001-6896-3282

**Shohei Chiashi** − Department of Mechanical Engineering, The University of Tokyo, Tokyo 113-8656, Japan; orcid.org/0000-0002-3813-0041

**Dmitry I. Levshov** − Theory and Spectroscopy of Molecules and Materials, Department of Chemistry, University of Antwerp, Antwerp 2610, Belgium; orcid.org/0000-0002-2249-7172

**Wim Wenseleers** − Nanostructured and Organic Optical and Electronic Materials, Department of Physics, University of Antwerp, Antwerp 2610, Belgium; orcid.org/0000-0002-3509-0945

**Keigo Otsuka** − Department of Mechanical Engineering, The University of Tokyo, Tokyo 113-8656, Japan; orcid.org/0000-0002-6694-0738

**Yan Li** − Center Beijing National Laboratory for Molecular Science, Key Laboratory for the Physics and Chemistry of Nanodevices, State Key Laboratory of Rare Earth Materials Chemistry and Applications, College of Chemistry and Molecular Engineering, Peking University, Beijing, 100871, China; orcid.org/0000-0002-3828-8340

**Christophe Bichara** − CINaM, CNRS - Aix Marseille University, Marseille 13288, France; orcid.org/0000-0003-4075-4563

**Sofie Cambré** − Theory and Spectroscopy of Molecules and Materials, Department of Physics, University of Antwerp, Antwerp 2610, Belgium; orcid.org/0000-0001-7471-7678

**Author Contributions**

Qingmei Hu conducted synthesis, optical characterization and analysis, TEM characterization and analysis, and data interpretation. Daniel Hedman and Daisuke Asa performed machine learning force field development, molecular dynamics simulation and analysis of results. Christophe Bichara provided theoretical insights for interpreting the MD simulations. Kaoru Hisama, Gunjan Auti, Hirofumi Daiguji and Keigo Otsuka partly joined the discussion of MD simulation and revised the manuscript. Wanyu Dai and Yongjia Zheng performed the STEM characterization. Yixi Yao assisted in the optical characterization of polymer-dispersed samples. Yan Li provided partial research suggestions. Sofie Cambré, Wim Wenseleers, Aina Fitó-Parera, and Dmitry I. Levshov contributed constructive suggestions to the absorption fitting procedure. Shigeo Maruyama, Ya Feng, and Rong Xiang supervised the overall research. The manuscript was written by Qingmei Hu, and all authors have approved the final version of the manuscript.

**Funding Sources**

This work was financially supported by the Japan Society for the Promotion of Science (JSPS) KAKENHI under Grant Numbers JP24KJ0751, JP23H00174, JP23H05443, and JP21KK0087, as well as by JST CREST under Grant Number JPMJCR20B5, Japan. Additionally, this work was supported by a JSPS-FWO Bilateral Joint Research Project (Grant Numbers JSPS JPJSBP120212301 and FWO VS08521N). Y.F. acknowledges support from the National Natural Science Foundation of China (No. 524766164), The Talent Development Program for Liaoning Province (No. XLYC2403036), as well as Science and Technology Innovation Fund of Dalian, International Science and Technology Cooperation (No. 2024JJ12RC035). A.F.P. acknowledges the Research Foundation Flanders (FWO) through a personal PhD fellowship (1178324N). R.X.

acknowledges the support by the National Key R&D Program of China (2024YFA1409600, 2023YFE0101300) from the Ministry of Science and Technology of China and research fund from Zhejiang province (2022R01001). D.H. acknowledge financial support from the Institute for Basic Science Korea (IBS-R019-D1).

**Notes**

The authors declare no competing financial interest.

## ACKNOWLEDGMENT

D. H would like to acknowledge the computational resources provided by the Institute for Basic Science (Korea) at the compute clusters Simulator (CMCM, Ulsan).

## ABBREVIATIONS

SWCNT, Single-Walled Carbon Nanotube; Iron, Fe; Cobalt, Co; CVD, Chemical Vapor Deposition; MD, Molecular dynamics; MLFF: Machine-Learned Force Field; NEP: Neuroevolution Potential; CNT, Carbon Nanotube; NPs, Nanoparticles; MWCNT, Multi-Walled Carbon Nanotube; TEM, Transmission electron microscopy; HAADF-STEM, High-angle annular dark-field scanning TEM; EDS: Energy-Dispersive X-ray Spectroscopy; SEM, Scanning electron microscopy; PLE, Photoluminescence-excitation; DOC, Sodium Deoxycholate; $D_2O$, Deuterium oxide; 2D PLE, Two-dimensional photoluminescence-excitation; FWHM, Full width at half maximum; ACCVD, Alcohol Catalytic Chemical Vapor Deposition; TGA, Thermogravimetric analysis.

## REFERENCES

(1) Ado, J.; Dresselhaus, G.; Dresselhaus, M. S. Carbon Nanotubes: Advanced Topics in the Synthesis, Structure, Properties and Applications. *Springer* **2008**.

(2) Yang, F.; Wang, M.; Zhang, D.; Yang, J.; Zheng, M.; Li, Y. Chirality Pure Carbon Nanotubes: Growth, Sorting, and Characterization. *Chem. Rev.* **2020**, 120 (5), 2693-2758.

(3) Maruyama, S.; Arnold, M. S.; Krupke, R.; Peng, L.-M. Physics and Applications of Nanotubes. *J. Appl. Phys.* **2022**, 131, 080401.

(4) Zhao, X.; Sun, S.; Yang, F.; Li, Y. Atomic-scale Evidence of Catalyst Evolution for the Structure-controlled Growth of Single-walled Carbon Nanotubes. *Acc. Chem. Res.* **2022**, 23, 3334-3344.

(5) Park, S.-H.; King, P. J.; Tian, R.; Boland, C. S.; Coelho, J.; Zhang, C. J.; McBean, P.; McEvoy, N.; Kremer, M. P.; Daly, D.; Coleman, J. N.; Nicolosi, V. High Areal Capacity Battery Electrodes Enabled by Segregated Nanotube Networks. *Nat. Energy* **2019**, 4, 560–567.

(6) He, Z.; Xiao, Z.; Yue, H.; Jiang, Y.; Zhao, M.; Zhu, Y.; Yu, C.; Zhu, Z.; Lu, F.; Jiang, H.; Zhang, C.; Wei, F. Single-walled Carbon Nanotube Film as an Efficient Conductive Network for Si-based Anodes. *Adv. Funct. Mater.* **2023**, 33 (26), 2300094.

(7) Lijun, L.; Han, J.; Xu, L.; Zhou, J.; Zhao, C.; Ding, S.; Shi, H.; Xiao, M.; Ding, L.; Ma, Z.; Jin, C.; Zhang, Z.; Peng, L-M. Aligned, High-density Semiconducting Carbon Nanotube Arrays for High-performance Electronics. *Science* **2020**, 368 (6493), 850-856.

(8) Foradori, S. M.; Jinkins, K. R.; Arnold, M. S. Link Among Array Nonuniformity, Threshold Voltage, and Subthreshold Swing Degradation in Aligned Array Carbon Nanotube Field Effect Transistors. *J. Appl. Phys.* **2020**, 128, 234301.

(9) Franklin, A. D. The Road to Carbon Nanotube Transistors. *Nature* **2013**, 498, 443.

(10) Hills, G.; Lau, C.; Wright, A.; Fuller, S.; Bishop, M. D.; Srimani, T.; Kanhaiya, P.; Ho, R.; Amer, A.; Stein, Y.; Murphy, D.; Arvind; Chandrakasan, A.; Shulaker, M. M. Modern Microprocessor Built from Complementary Carbon Nanotube Transistors. *Nature* **2019**, 572, 595.

(11) Feng, Y.; Wang, M.; Zhang, D.; Yang, J.; Zheng, M.; Li, Y. Chirality Pure Carbon Nanotubes: Growth, Sorting, and Characterization. *Chem. Rev.* **2020**, 120 (5), 2693-2758.

(12) Jing, K.; Cassell, A. M.; Dai, H. Chemical Vapor Deposition of Methane for Single-walled Carbon Nanotubes. *Chem. Phys. Lett.* **1998**, 292 (4-6), 567-574.

(13) Hongjie, D.; Rinzler, A. G.; Nikolaev, P.; Thess, A.; Colbert, D. T.; Smalley, R. E. Single-wall Nanotubes Produced by Metal-catalyzed Disproportionation of Carbon Monoxide. *Chem. Phys. Lett.* **1996**, 260 (3-4), 471-475.

(14) Maruyama, S.; Kojima, R.; Miyauchi, Y.; Chiashi, S.; Kohno, M. Low-temperature Synthesis of High-purity Single-walled Carbon Nanotubes from Alcohol. *Chem. Phys. Lett.* **2002**, 360 (3-4), 229-234.

(15) Wang, B.; Poa, C. H. P.; Wei, L.; Li, L.-J.; Yang, Y.; Chen, Y. (n,m) Selectivity of Single-Walled Carbon Nanotubes by Different Carbon Precursors on Co−Mo Catalysts. *J. Am. Chem. Soc.* **2007**, 129 (29), 9014–9019.

(16) Yang, F.; Wang, X.; Zhang, D.; Yang, J.; Luo, D.; Xu, Z.; Wei, J.; Wang, J. Q.; Xu, Z.; Peng, F.; Li, X.; Li, R.; Li, Y.; Li, M.; Bai, X.; Ding, F.; Li, Y. Chirality-specific Growth of Single-walled Carbon Nanotubes on Solid Alloy Catalysts. *Nature* **2014**, 510 (7506), 522-524.

(17) Zhang, S.; Kang, L.; Wang, X.; Tong, L.; Yang, L.; Wang, Z.; Qi, K.; Deng, S.; Li, Q.; Bai, X.; Ding, F.; Zhang, J. Arrays of Horizontal Carbon Nanotubes of Controlled Chirality Grown Using Designed Catalysts. *Nature* **2017**, 543, 234–238.

(18) Kitiyanan, B.; Alvarez, W. E.; Harwell, J. H.; Resasco, D. E. Controlled Production of Single-Wall Carbon Nanotubes by Catalytic Decomposition of CO on Bimetallic Co–Mo Catalysts. *Chem. Phys. Lett.* **2000**, 317, 497-503.

(19) Hu, M.; Murakami, Y.; Ogura, M.; Maruyama, S.; Okubo, T. Morphology and Chemical State of Co-Mo Catalysts for Growth of Single-walled Carbon Nanotubes Vertically Aligned on Quartz Substrates. *J. Catal.* **2004**, 225 (1), 230-239.

(20) Hou, B.; Wu, C.; Inoue, T.; Chiashi, S.; Xiang, R.; Maruyama, S. Extended Alcohol Catalytic Chemical Vapor Deposition for Efficient Growth of Single-walled Carbon Nanotubes Thinner than (6,5). *Carbon* **2017**, 119, 502-510.

(21) Miyauchi, Y.; Chiashi, S.; Murakami, Y.; Hayashida, Y.; Maruyama, S. Fluorescence Spectroscopy of Single-walled Carbon Nanotubes Synthesized from Alcohol. *Chem. Phys. Lett.* **2004**, 387 (1-3), 198-203.

(22) Murakami, T.; Mitikami, K.; Ishigaki, S.; Matsumoto, K.; Nishio, K.; Isshiki, T.; Harima, H. Catalytic Mechanism of a Fe–Co Bimetallic System for Efficient Growth of Single-walled Carbon Nanotubes on Si/$SiO_2$ Substrates. *J. Appl. Phys.* **2006**, 100 (094303).

(23) Chiang, W.-H.; Sankaran, R. M. The Influence of Bimetallic Catalyst Composition on Single-Walled Carbon Nanotube Yield. *Carbon* **2012**, 50 (3), 1044-1050.

(24) Xiang, R.; Einarsson, E.; Murakami, Y.; Shiomi, J.; Chiashi, S.; Tang, Z.; Maruyama, S. Diameter Modulation of Vertically Aligned Single-Walled Carbon Nanotubes. *ACS Nano* **2012**, 6 (8), 7472–7479.

(25) An, H.; Kumamoto, A.; Xiang, R.; Inoue, T.; Otsuka, K.; Chiashi, S.; Bichara, C.; Loiseau, A.; Li, Y.; lkuhara, Y.; Maruyama, S. Atomic-scale Structural Identification and Evolution of Co-W-C Ternary SWCNT Catalytic Nanoparticles High-resolution STEM Imaging on $SiO_2$. *Sci. Adv.* **2019**, 5 (5), eaat9459.

(26) Cui, K.; Kumamoto, A.; Xiang, R.; An, H.; Wang, B.; Inoue, T.; Chiashi, S.; Ikuhara, Y.; Maruyama, S. Synthesis of Sub Nanometer-Diameter Vertically Aligned Single-Walled Carbon Nanotubes with Copper-Anchored Cobalt Catalysts. *Nanoscale* **2016**, 8 (3), 1608-1617.

(27) Shiina, S.; Murohashi, T.; Ishibashi, K.; He, X.; Koretsune, T.; Liu, Z.; Terashima, W.; Kato, Y. K.; Inoue, K.; Saito, M.; Ikuhara, Y.; Kato, T. Synthesis of Ultrahigh-Purity (6,5) Carbon Nanotubes Using a Trimetallic Catalyst. *ACS Nano* **2024**, 18 (35), 23979-23990.

(28) Zhang, F.; Zhang, L.; Jiang, H.; Li, X.; Liu, F.; Ji, Z.-H.; Hou, P.-X.; Guo, S.; Cheng, H.-M.; Kauppinen, E. I.; Liu, C.; Ding, F. Growth of High-density Single-wall Carbon Nanotubes with a Uniform Structure Using a CoRu Catalyst. *Carbon* **2023**, 209.

(29) Everhart, B. M.; Rao, R.; Nikolaev, P.; Liu, T.-W.; Gómez-Gualdrón, D. A.; Maruyama, B.; Amama, P. B. High-Throughput Experimentation for Selective Growth of Small-Diameter Single-Wall Carbon Nanotubes Using Ru-Promoted Co Catalysts. *Chem. Mater.* **2022**, 34 (10), 4548–4559.

(30) Page, A. J.; Villamanca, D. M.; Amama, P. B.; McLean, B. Electronic Structure of Bimetallic CoRu Catalysts Modulates the Early Stages of SWCNT Nucleation. *Nanoscale* **2026**.

(31) Maruyama, S. Super Mackay Cluster $Fe_{42}Co_{13}$ for Chirality Selective Growth of Single-Walled Carbon Nanotubes. *NT23*, *Arcachon* **2023**.

(32) Liu, M.; An, H.; Kumamoto, A.; Inoue, T.; Chiashi, S.; Xiang, R.; Maruyama, S. Efficient Growth of Vertically-aligned Single-walled Carbon Nanotubes Combining Two Unfavorable Synthesis Conditions. *Carbon* **2019**, 146, 413-419.

(33) Xiang, R.; Maruyama, S. Revisiting Behavior of Monometallic Catalysts in Chemical Vapor Deposition Synthesis of Single-Walled Carbon Nanotubes. *R Soc Open Sci* **2018**, 5 (8), 180345.

(34) Tairan Wang; Jianyu Hu; Runhai Ouyang; Yutao Wang; Yi Huang; Sulei Hu; Li, W.-X. Nature of Metal-support Interaction for Metal Catalysts on Oxide Supports. *Science* **2024**, 386, 915–920.

(35) Lin, M.; Ying, J. P.; Boothroyd, T.; Loh, K. P.; Tok, E. S.; Foo, Y.-L. Direct Observation of Single-Walled Carbon Nanotube Growth at the Atomistic Scale. Nano Letters 2006, 6 (3), 449–452.

(36) Calizzi, M.; Mutschler, R.; Patelli, N.; Migliori, A.; Zhao, K.; Pasquini, L.; Zuttel, A. CO(2) Hydrogenation Over Unsupported Fe-Co Nanoalloy Catalysts. *Nanomaterials* **2020**, 10 (7).

(37) Dupuis, V.; Khadra, G.; Montejano-Carrizales, J. M.; Tournus, F.; Aguilera-Granja, F.; Tamion, A. Mass-Selected FeCo Clusters Embedded in a Carbon Matrix as Benchmark Nanocatalysts. *ACS Applied Nano Materials* **2019**, 2 (5), 2864-2872.

(38) Xiang, R.; Wu, T. Z.; Einarsson, E.; Suzuki, Y.; Murakami, Y.; Shiomi, J.; Maruyama, S. High-precision Selective Deposition of Catalyst for Facile Localized Growth of Single-walled Carbon Nanotubes. *J. Am. Chem. Soc.* **2009**, 131, 10344-10345.

(39) Murakami, Y.; Chiashi, S.; Miyauchi, Y.; Hu, M.; Ogura, M.; Okubo, T.; Maruyama, S. Growth of Vertically Aligned Single-walled Carbon Nanotube Films on Quartz Substrates and Their Optical Anisotropy. *Chem. Phys. Lett.* **2004**, 385 (3-4), 298-303.

(40) Wenseleers, W.; Vlasov, I. I.; Goovaerts, E.; Obraztsova, E. D.; Lobach, A. S.; Bouwen, A. Efficient Isolation and Solubilization of Pristine Single-walled Nanotubes in Bile Salt Micelles. *Adv. Funct. Mater.* **2004**, *14* (11), 1105-1112.

(41) Tian, Y.; Jiang, H.; Anoshkin, I. V.; Kauppinen, L. J. I.; Mustonen, K.; Nasibulin, A. G.; Kauppinen, E. I. A Reference Material of Single-walled Carbon Nanotubes: Quantitative Chirality Assessment Using Optical Absorption Spectroscopy. *RSC Adv.* **2015**, 5 (125), 102974-102980.

(42) Pfohl, M.; Tune, D. D.; Graf, A.; Zaumseil, J.; Krupke, R.; Flavel, B. S. Fitting Single-Walled Carbon Nanotube Optical Spectra. *ACS Omega* **2017**, 2 (3), 1163-1171.

(43) Tsyboulski, D. A.; Rocha, J.-D. R.; Bachilo, S. M.; Cognet, L.; Weisman, R. B. Structure-Dependent Fluorescence Efficiencies of Individual Single-Walled Carbon Nanotubes. *Nano Lett.* **2007**, 7 (10), 3080-3085.

(44) Campo, J.; Cambré, S.; Botka, B.; Obrzut, J.; Wenseleers, W.; Fagan, J. A. Optical Property Tuning of Single-Wall Carbon Nanotubes by Endohedral Encapsulation of a Wide Variety of Dielectric Molecules. *ACS Nano* **2021**, 15 (2), 2301-2317.

(45) Cambré, S.; Van Werveke, W.; De Clercq, M.; Erkens, M.; Martinati, M.; Wenseleers, W. Quantitative 2D Fitting of Fluorescence-excitation Maps: Excitation Lineshape of Single-wall Carbon Nanotubes. *Nanoscale Horiz.* **2025**.

(46) Fan, Z.; Chen, W.; Vierimaa, V.; Harju, A. Efficient Molecular Dynamics Simulations with Many-body Potentials on Graphics Processing Units. *Comput. Phys. Commun.* **2017**, 218, 10-16.

(47) Bussi, G.; Donadio, D.; Parrinello, M. Canonical Sampling Through Velocity Rescaling. *J Chem Phys* **2007**, 126 (1), 014101.

(48) Fan, Z.; Zeng, Z.; Zhang, C.; Wang, Y.; Song, K.; Dong, H.; Chen, Y.; Ala-Nissila, T. Neuroevolution Machine Learning Potentials: Combining High Accuracy and Low Cost in Atomistic Simulations and Application to Heat Transport. *Phys. Rev. B* **2021**, 104 (10).

(49) Fan, Z. Improving the Accuracy of the Neuroevolution Machine Learning Potential for Multi-Component Systems. *J Phys Condens Matter* **2022**, 34 (12).

(50) Fan, Z.; Wang, Y.; Ying, P.; Song, K.; Wang, J.; Wang, Y.; Zeng, Z.; Xu, K.; Lindgren, E.; Rahm, J. M. GPUMD: A Package for Constructing Accurate Machine-learned Potentials and Performing Highly Efficient Atomistic Simulations. *J Chem Phys* **2022**, 157 (11), 114801.

(51) Song, K.; Zhao, R.; Liu, J.; Wang, Y.; Lindgren, E.; Wang, Y.; Chen, S.; Xu, K.; Liang, T.; Ying, P. General-purpose Machine-learned Potential for 16 Elemental Metals and Their Alloys. *Nat. Commun.* **2024**, 15 (1).

(52) Stukowski, A. Visualization and Analysis of Atomistic Simulation Data with OVITO–the Open Visualization Tool. *Modell. Simul. Mater. Sci. Eng.* **2010**, 18 (1).

**Supplementary Information**

# Structures of iron and cobalt bimetallic clusters for optimized chemical vapor deposition growth of single-walled carbon nanotubes

*Qingmei Hu[1], Daniel Hedman[2*], Ya Feng[3*], Wanyu Dai[1], Daisuke Asa[1], Aina Fitó-Parera[4], Yixi Yao[5], Yongjia Zheng[6], Kaoru Hisama[7], Gunjan Auti[1], Hirofumi Daiguji[1], Shohei Chiashi[1], Dmitry Levshov[4], Wim Wenseleers[9], Keigo Otsuka[1], Yan Li[5], Christophe Bichara[8], Sofie Cambré[4], Rong Xiang[6*], Shigeo Maruyama[1,6,10*]*

## 1. Supplementary figures for the discussion in the main text

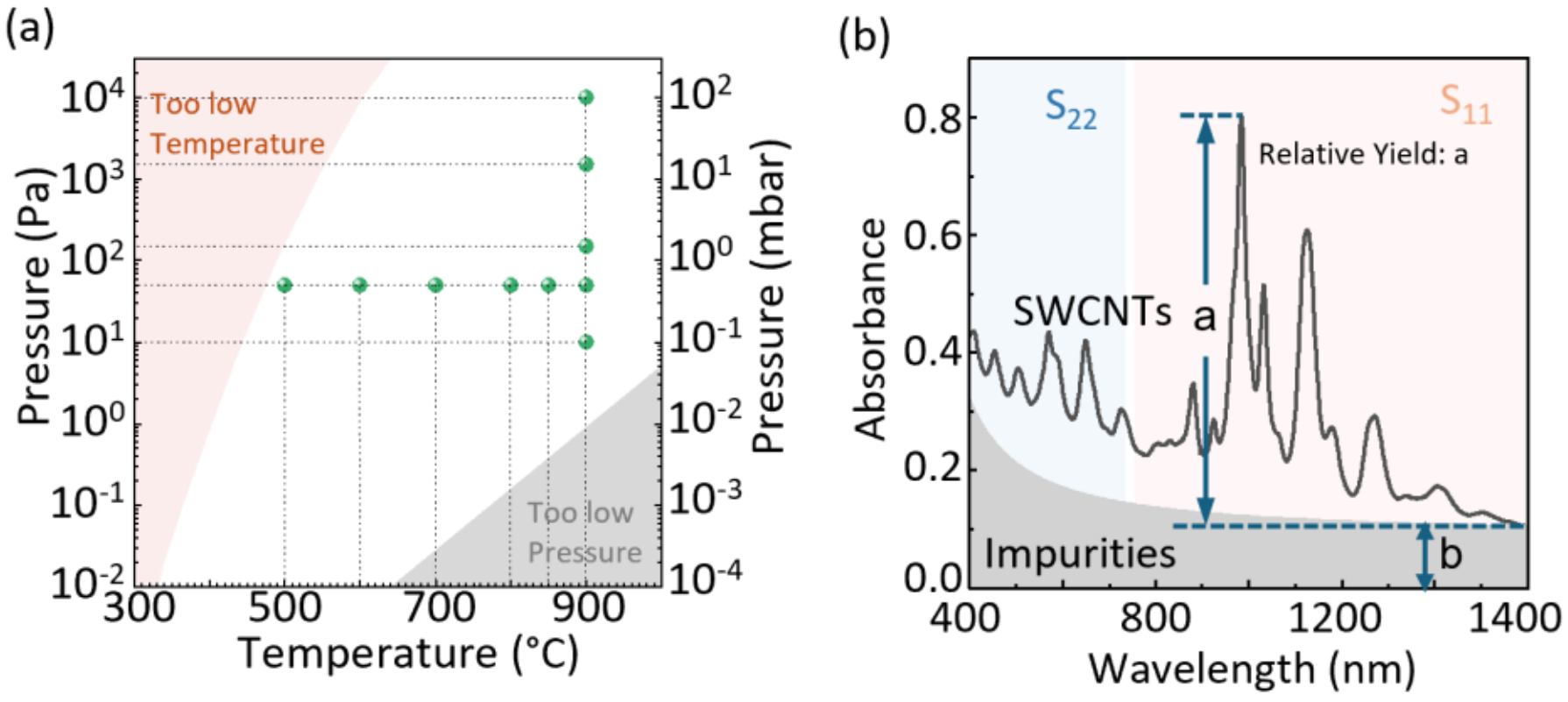

**Figure S1.** Study of SWCNTs' growth through ACCVD. (a) The broad operational window of ACCVD.[1] (b) A typical baseline subtraction of a UV-vis-NIR absorption spectrum using the Fano and Lorentzian function model developed by Tian *et al*.[2]

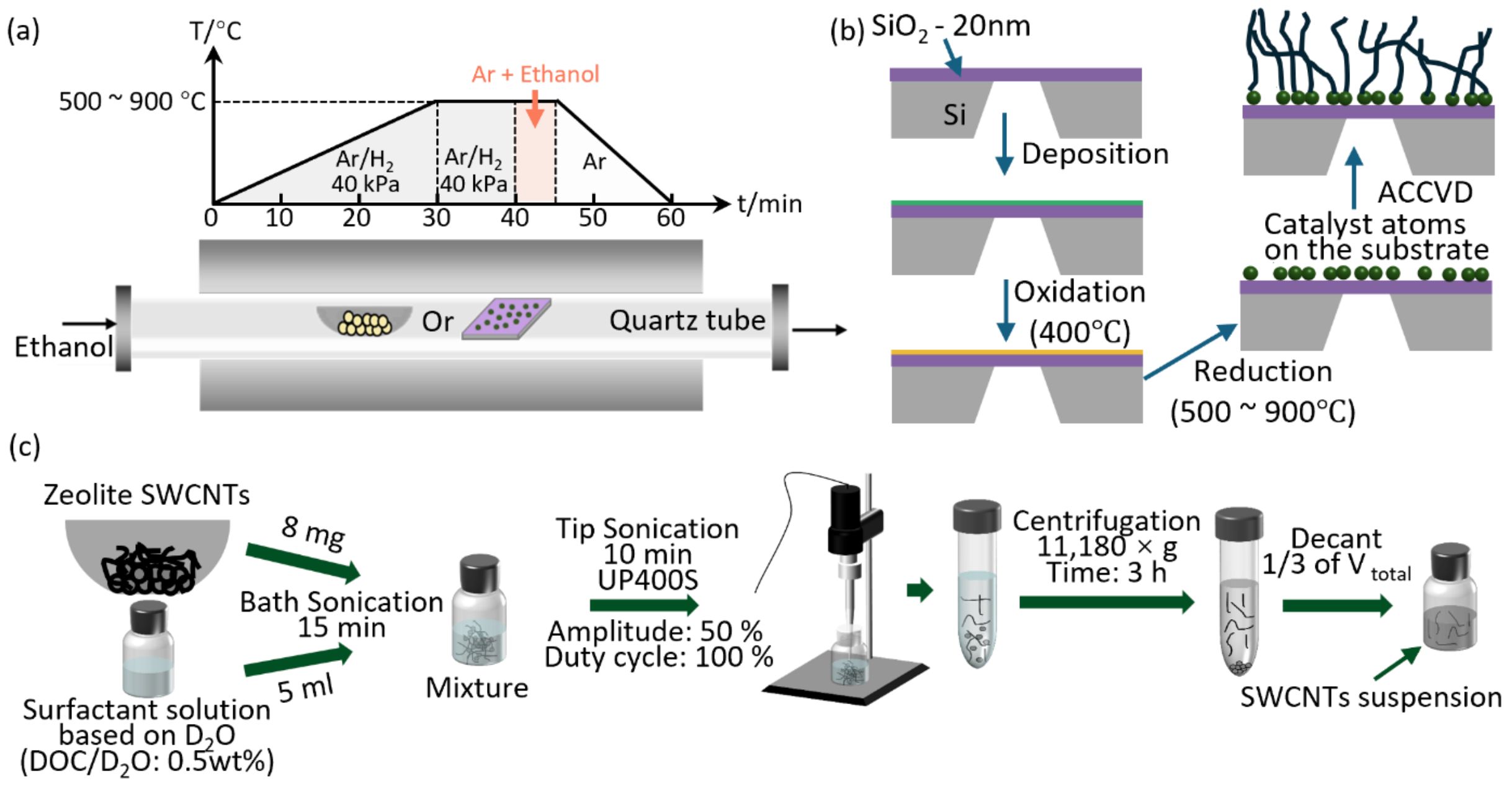

**Figure S2.** The sample preparation procedure. ACCVD for (a) zeolite-supported or (b) $SiO_2$/Si-supported SWCNT growth; (c) Dispersion processes for preparing suspensions of SWCNTs.

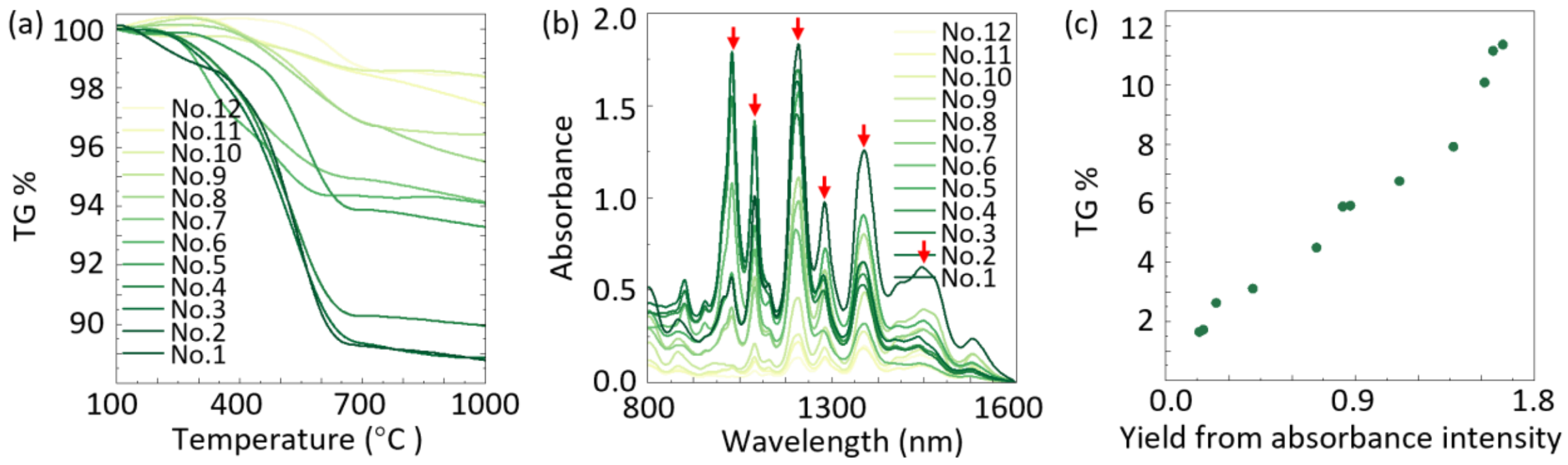

**Figure S3.** Verification of the relationship between TGA and absorbance. (a) TGA analysis and (b) Absorption spectra (using a 1cm pathlength cuvette.) of the samples obtained by 12 different combinations of growth parameters; The absorption peaks indicated by the red arrows are selected, and their intensities are summed to generate the data presented in Figure c. (c) The dependence of TGA weight loss with respect to absorbance, showing a very nice linear dependence, validating the use of absorbance measurements as a reliable indicator of SWCNT yield. Notably, zeolite represents the dominant component in the original sample, with SWCNTs comprising only around 10% of the total weight.

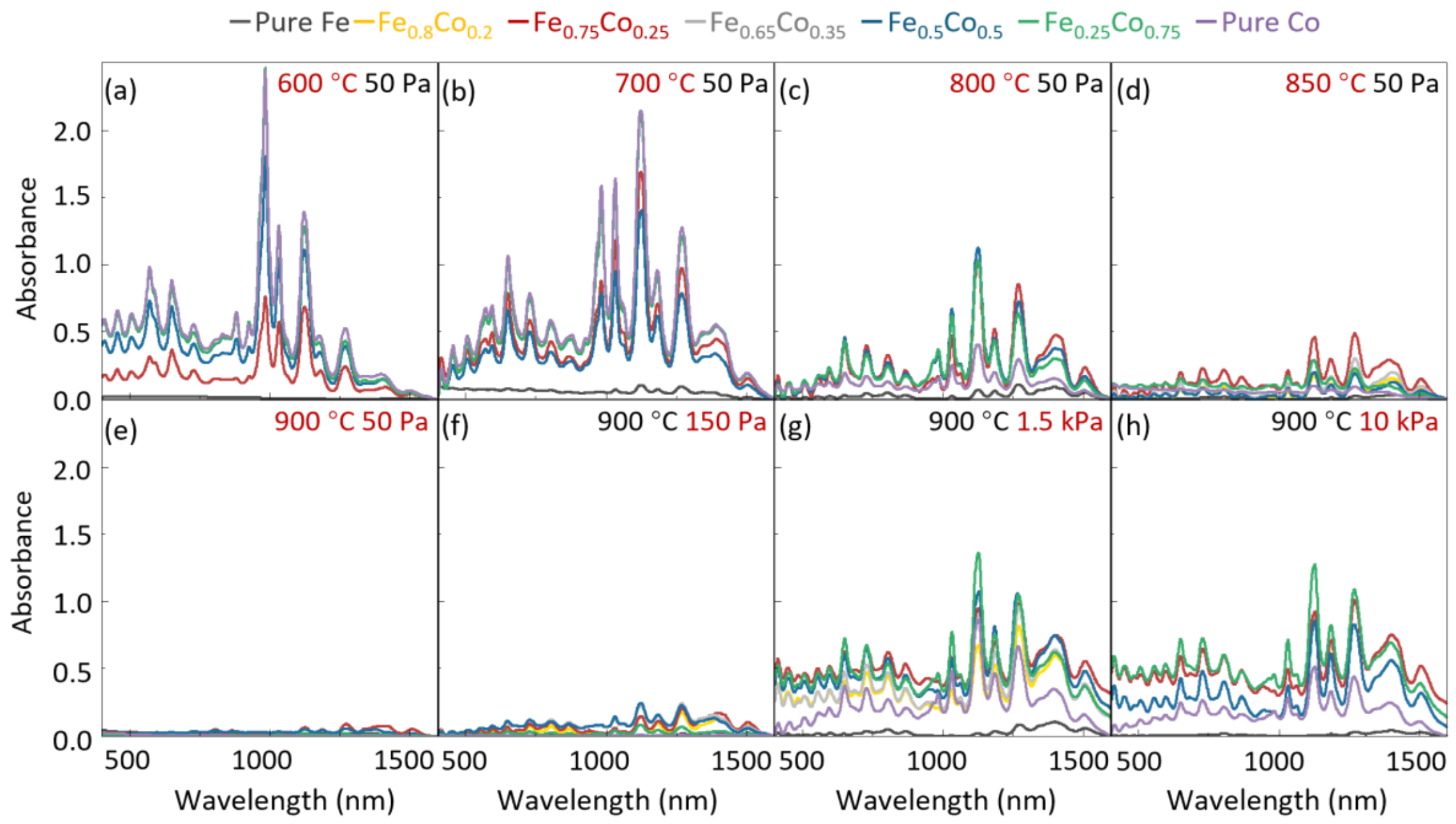

**Figure S4.** Absorption spectra for samples prepared under different growth conditions, with varying Fe-Co ratios at different temperatures or pressures.

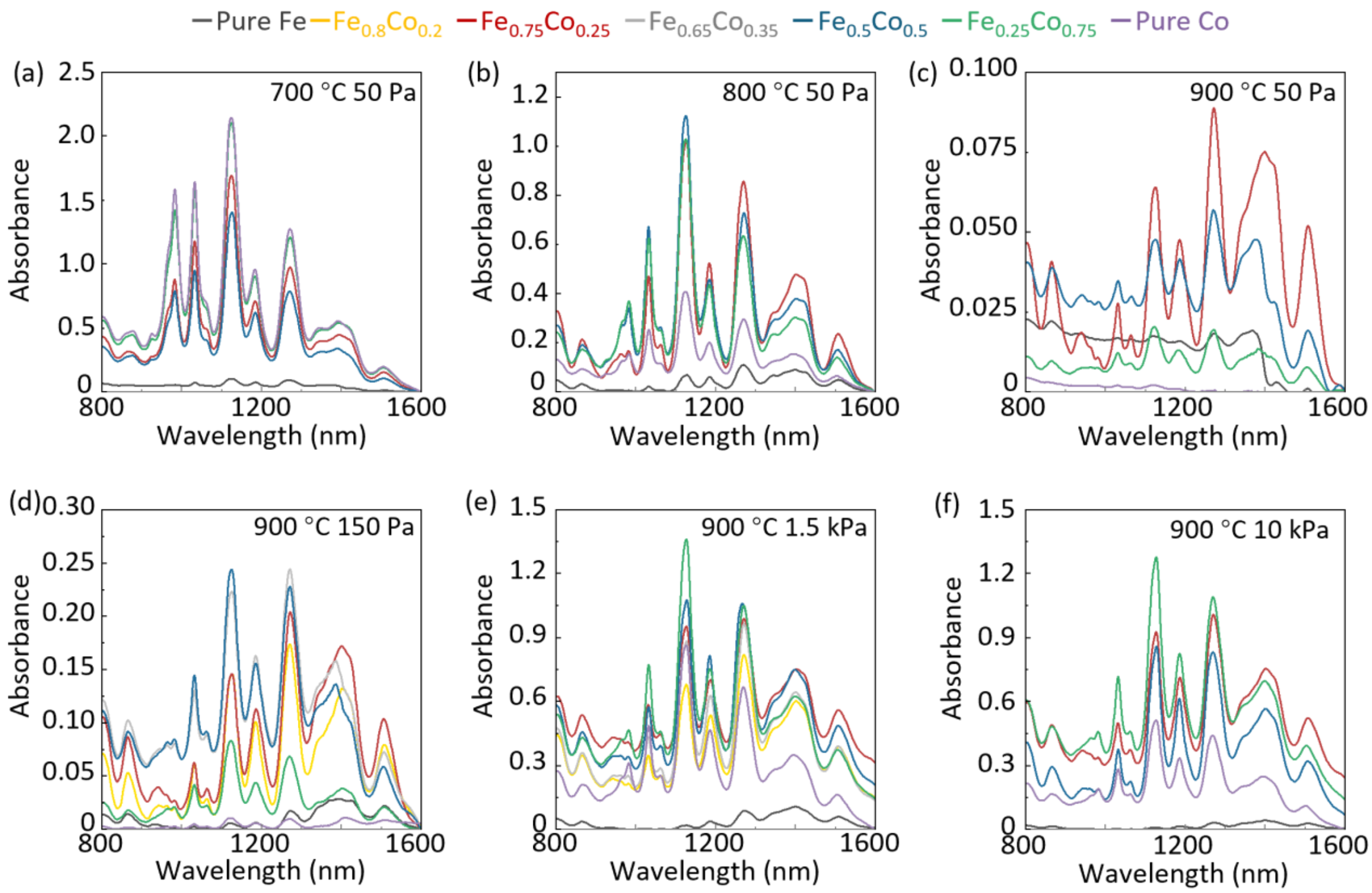

**Figure S5.** The absorption spectrum of samples grown at (a) 700 °C 50 Pa; (b) 800 °C 50 Pa; (c) 900 °C 50 Pa; (d) 900 °C 150 Pa; (e) 900 °C 1.5 kPa; (f) 900 °C 10 kPa, each with varying Fe-Co ratios. Note the different y-scales of the panels.

**Fig. S4** and **S5** present the absorbance spectra of two sets of SWCNT samples synthesized under varying growth conditions, including different Fe/Co atomic ratios in the catalysts, growth temperatures, and ethanol partial pressures. To enable a direct visual comparison across all conditions, each subplot in **Fig. S4** employs a unified coordinate range. As shown in **Fig. S4**, under 600 °C growth conditions, all catalysts except pure Fe exhibit high absorbance values, indicating generally high growth efficiency. In contrast, when the temperature increases to 900 °C while maintaining a low ethanol pressure of 50 Pa, absorbance values significantly decrease, with some samples falling below 0.1 (**Fig. S4e**). This suggests that such conditions approach the practical limit of growth, making reliable optical analysis infeasible. Conversely, at 900 °C, increasing the ethanol partial pressure leads to a marked enhancement in absorbance (**Fig. S4e–S4h**), highlighting the critical role of pressure in high-temperature growth. This trend aligns with previous studies showing that in ACCVD systems, both relatively low-temperature/low-pressure and relatively high-temperature/high-pressure conditions are favorable for SWCNT synthesis, while relatively high-temperature/low-pressure conditions represent a more stringent and less efficient growth regime. At low temperature high pressure is less favorable due to the catalyst deactivation.[1] To improve the clarity of individual spectra, the vertical axis range in each subplot was independently optimized in **Fig. S5**, allowing better visualization of variations in absorbance among samples.

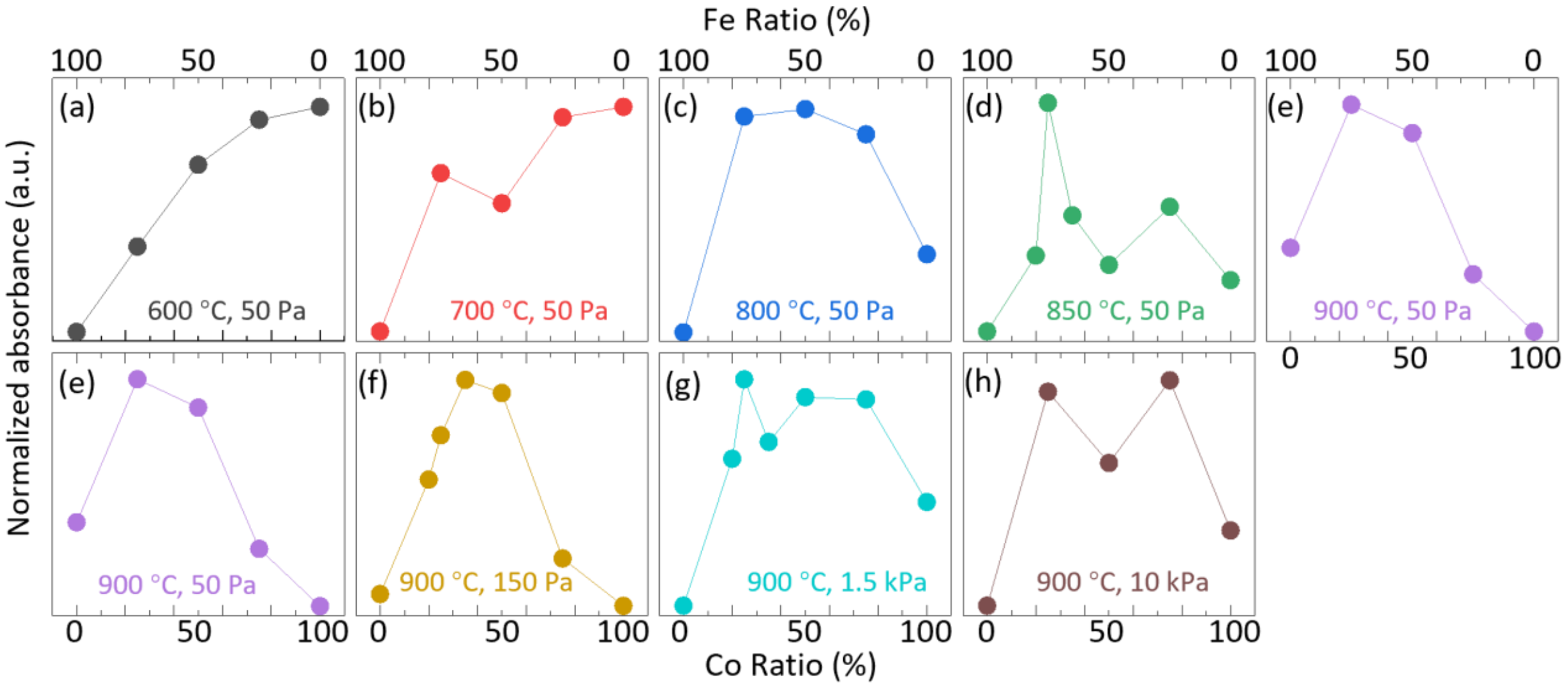

**Figure S6.** Effect of atomic ratio in bimetallic catalysts on the absorbance of SWCNTs grown under different ACCVD conditions: (a-e) different growth temperatures; and (e-h) different growth pressures.

As previously discussed, absorbance serves as a reliable indicator of SWCNT growth efficiency. To systematically evaluate the catalytic performance of Fe/Co bimetallic catalysts under various conditions, the absorbance spectra of individual samples were further processed (similar to **Fig. S3b**) to extract representative and strong absorbance values. These average values were then used to construct **Fig. S6**, which reveals the correlation between average absorbance and catalyst composition.

In the first series of experiments (**Fig. S6a–S6e**), conducted at a constant ethanol pressure of 50 Pa, the highest absorbance was observed for samples synthesized with pure Co catalysts at 600 °C, while those using pure Fe showed negligible activity. As the temperature increased, the optimal Fe/Co ratio gradually shifted toward higher Fe content. This behavior looks like Fe atoms began to participate in the catalytic process at elevated temperatures. However, this apparent synergistic effect between Fe and Co attributed to the fact that Fe is needed to stabilize the small nanoparticles at these higher temperatures, preventing them from coalescing. At 850 °C, the $Fe_{0.75}Co_{0.25}$ composition already showed the highest absorbance, which was further confirmed as the optimal catalyst at 900 °C, supporting the notion that increased Fe content enhances growth efficiency under these conditions.

In contrast, when the growth temperature was fixed at 900 °C and the ethanol partial pressure was increased (**Fig. S6e–S6h**), the optimal composition shifted toward higher Co content. At a pressure of 1.5 kPa, $Fe_{0.25}Co_{0.75}$ become as strong as $Fe_{0.75}Co_{0.25}$ emerged as the most effective catalyst. We speculate that at higher pressures, the stabilizing effect of the Fe is less critically needed, because at these higher pressures' growth is faster (because carbon is more abundant), and is completed before the nanoparticles have had the time to merge into larger nanoparticles. This would also explain the higher yields at high pressure (at high T) because at high temperature and low pressure, the growth ends prematurely because the nanoparticles merge into too large particles too rapidly. Based on the results shown in **Fig. S3** and **S6**, it can be inferred that under harsher growth conditions (relatively, high temperature, low pressure), Fe plays a more crucial role in sustaining growth, whereas under more favorable conditions (high temperature and high pressure or low temperature and low pressure, relatively), the catalytic activity is dominated by Co.

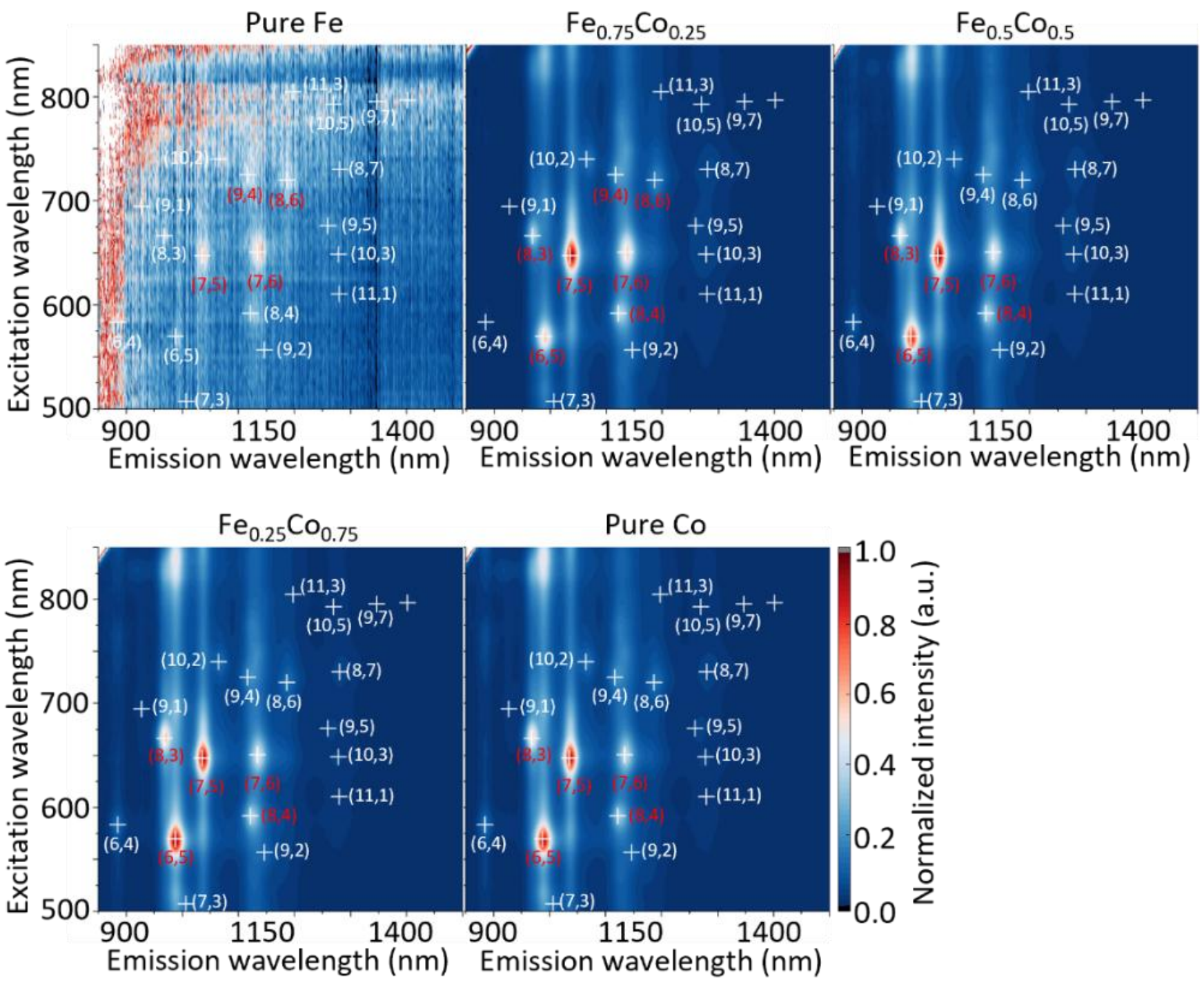

**Figure S7.** Normalized fluorescence-excitation (PLE) maps of dispersed solution samples of SWCNTs grown at 600 °C, 50Pa.

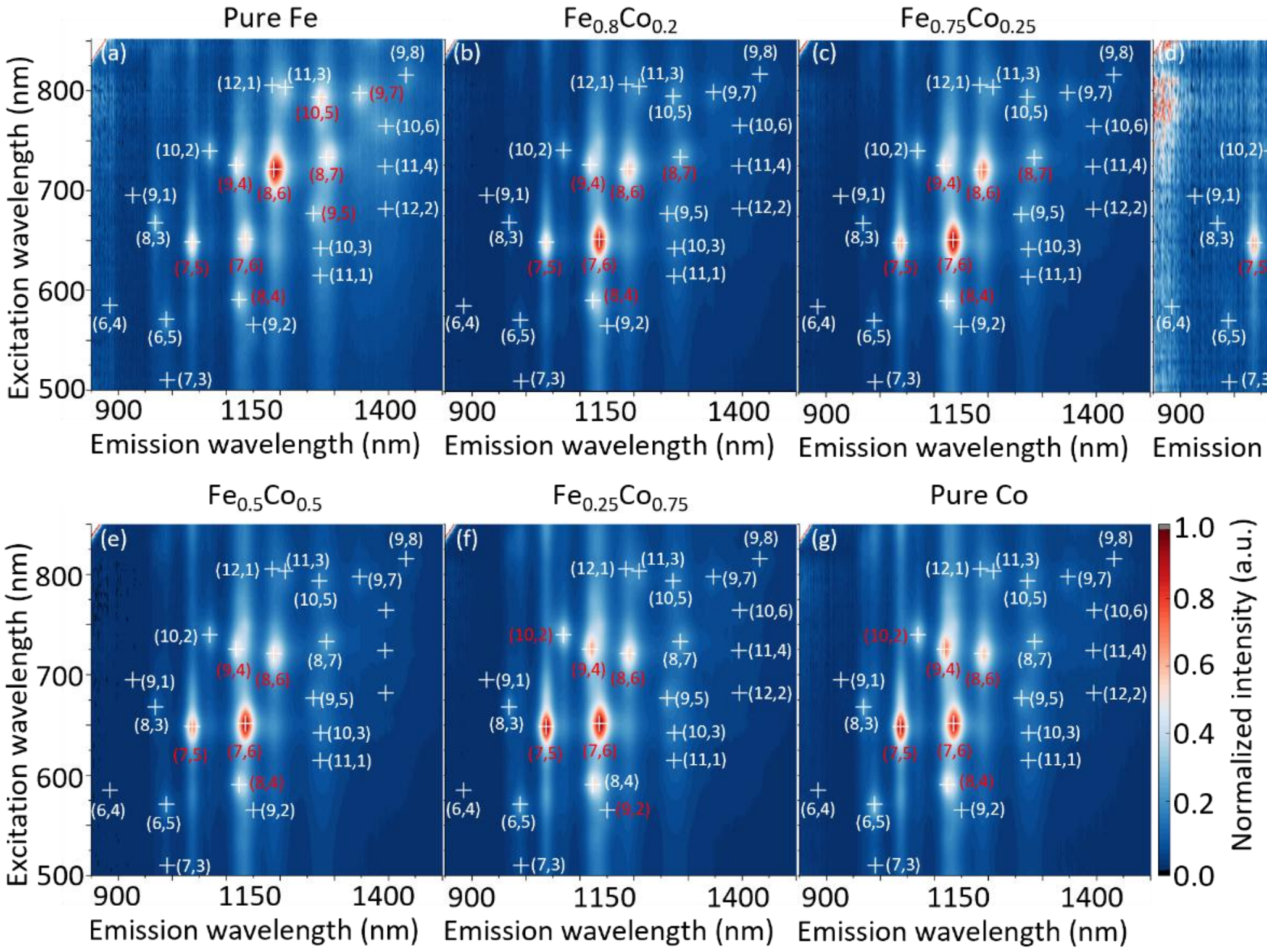

**Figure S8.** Normalized PLE maps of dispersed solution samples of SWCNTs grown at 850 °C, 50Pa.

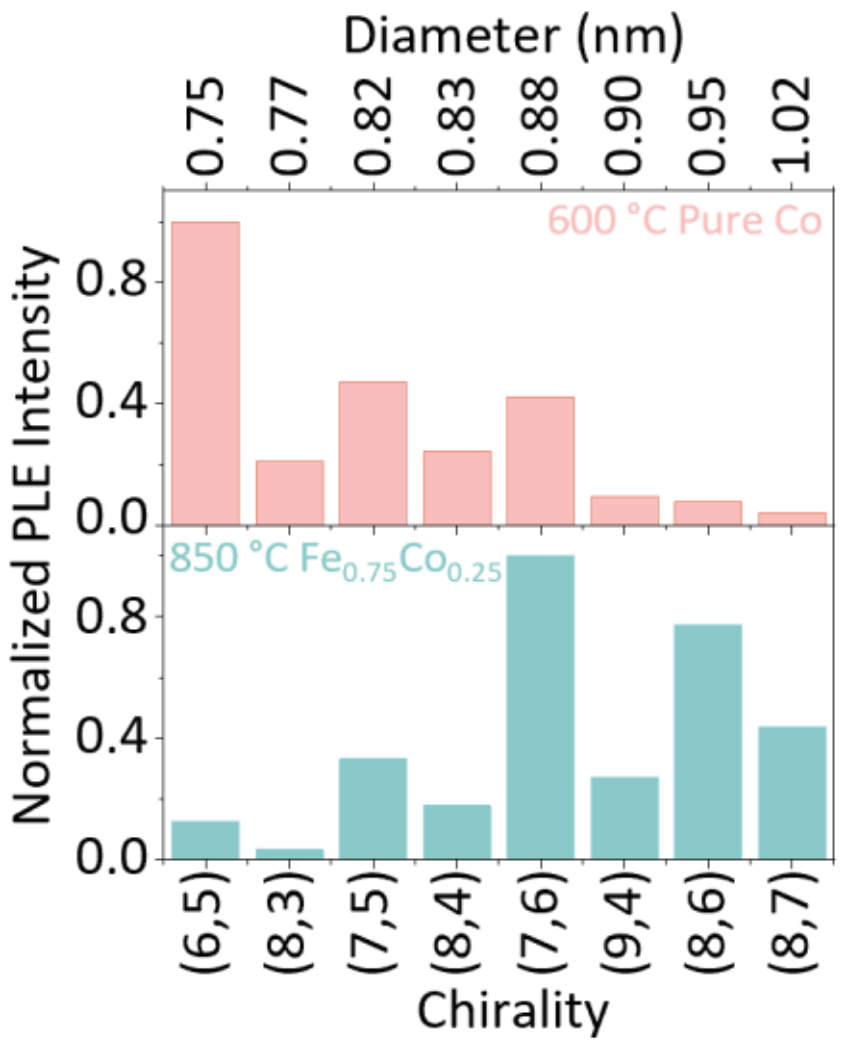

**Figure S9.** Normalized PLE intensity (from **Fig. S7** and **S8**) bar chart for SWCNTs grown using pure Co catalysts at 600 °C and $Fe_{0.75}Co_{0.25}$ catalysts at 850 °C. A clear shift in intensity, and thus chiral composition, towards larger diameter SWCNTs is observed between the two growth conditions.

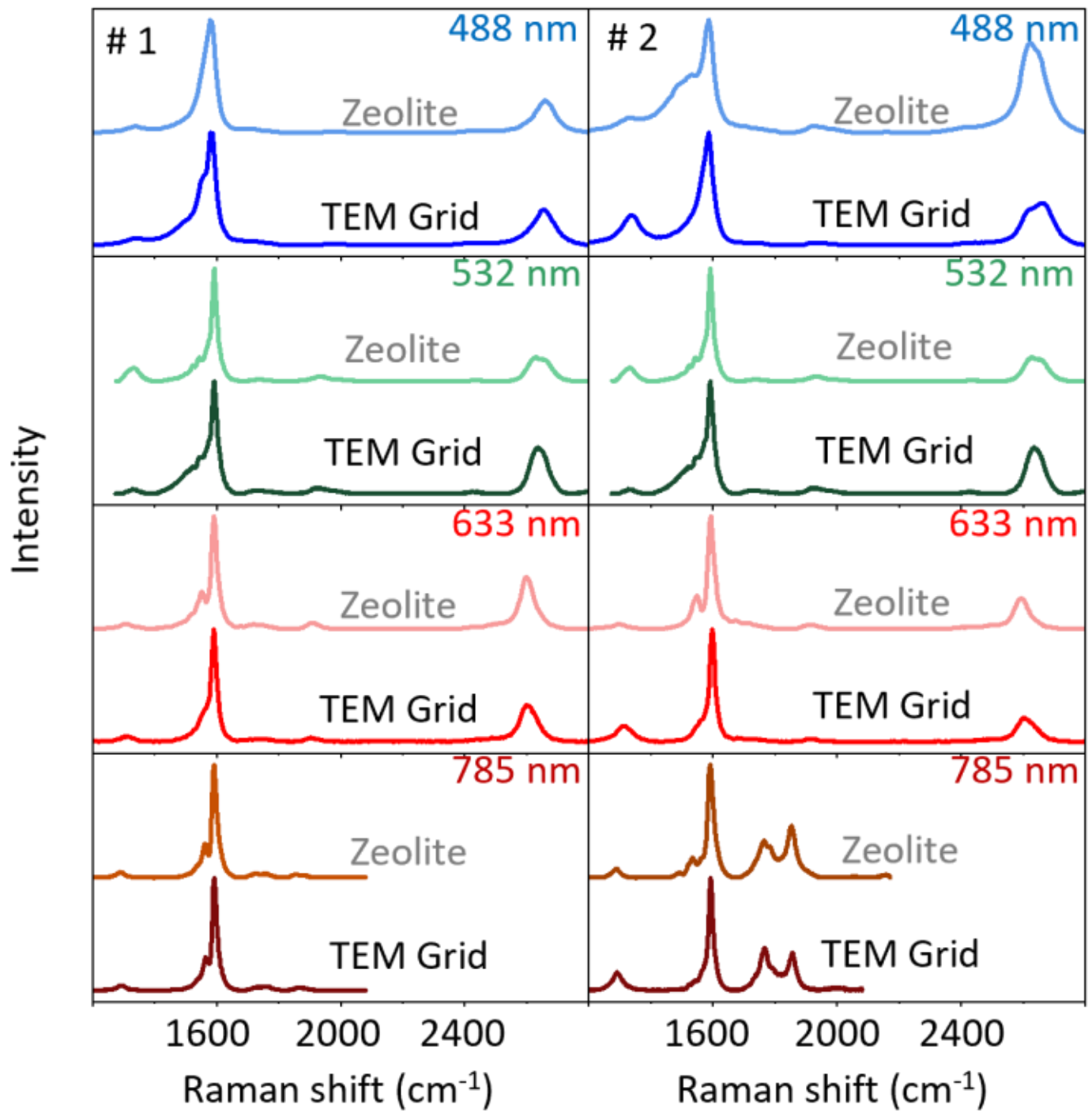

**Figure S10.** Raman spectra (in the range of D, G and 2D bands) for two sets of samples grown on zeolite and $SiO_2$/Si TEM grids. Here #1 is synthesized using $Fe_{0.5}Co_{0.5}$ at 800 °C, 1.2 kPa and #2 is synthesized using pure Co at 600 °C, 50 Pa, respectively.

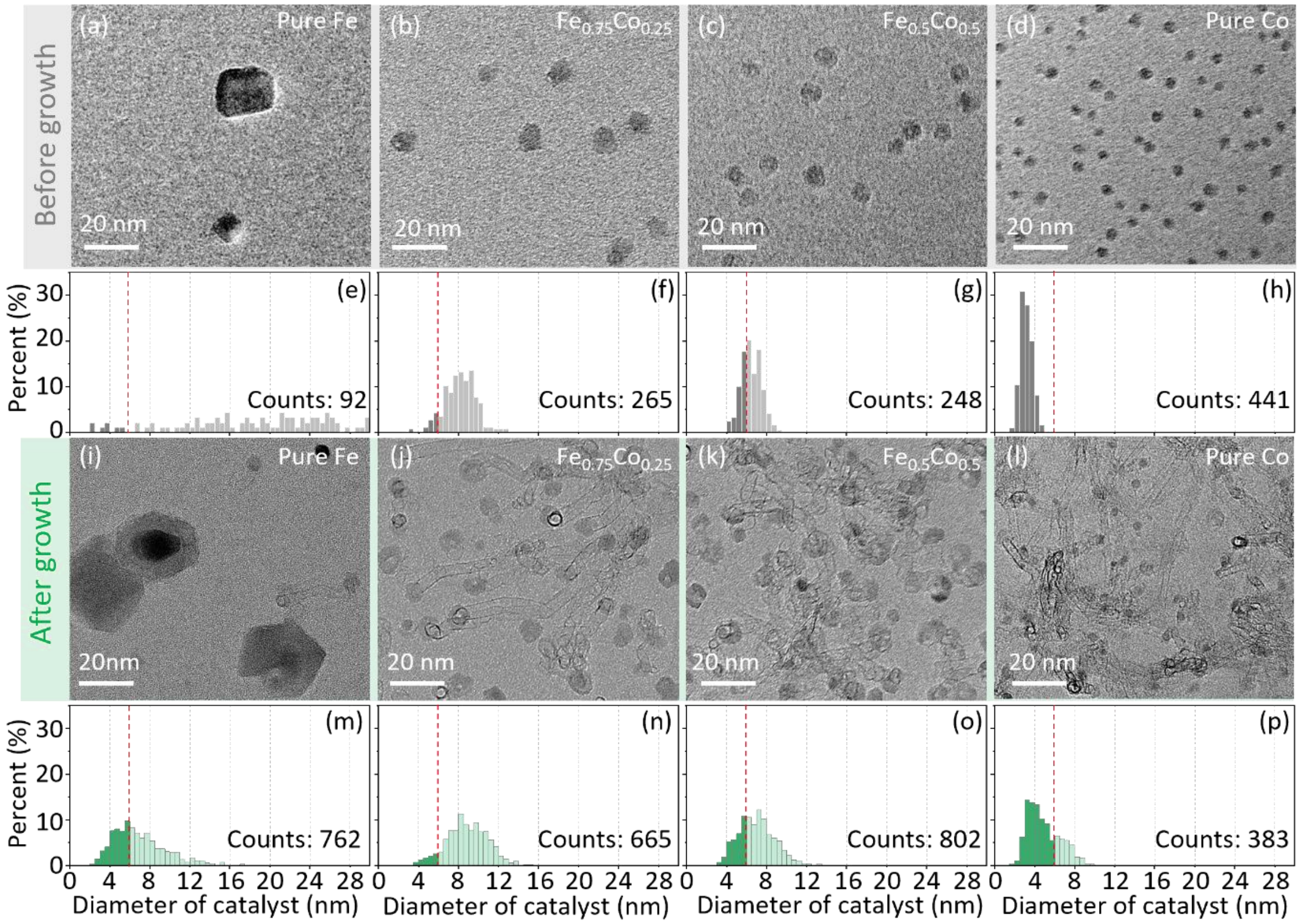

**Figure S11.** TEM characterization and analysis of samples grown at 600 °C and 50 Pa. TEM images of samples with different Fe-Co ratio before (a-d) and after (i-l) growth; Histogram of catalyst diameter for $Fe_xCo_{1-x}$ samples before (a-d) and after (i-l) growth.

In contrast to the data presented in **Fig. 4** in the main text at 850 °C, here we present the data at 600 °C. Before the ACCVD process, the catalysts of Pure Fe exhibited a broad particle size distribution, with the majority of particles ranging between 20 and 30 nm and only a small fraction measuring below 10 nm (**Fig. S11a**). As the Co ratio increased, the particle size distribution shifted toward smaller diameters, with pure Co catalysts showing the narrowest distribution, centered around 3 nm (**Fig. S11d**). Following the ACCVD process (**Fig. S11i – S11p**), all samples except those with pure Fe catalysts successfully induced SWCNT growth. Among the effective catalysts ($Fe_{0.75}Co_{0.25}$, $Fe_{0.5}Co_{0.5}$, and pure Co), the density of SWCNTs increased with increasing Co ratio, with pure Co catalysts demonstrating the highest SWCNT density. Notably, the post-growth catalyst particle sizes aligned with the pre-growth trends, reinforcing the critical influence of catalyst composition, namely $Fe_xCo_{1-x}$ ratio, and consequently particle size, on catalytic performance under these ACCVD conditions.

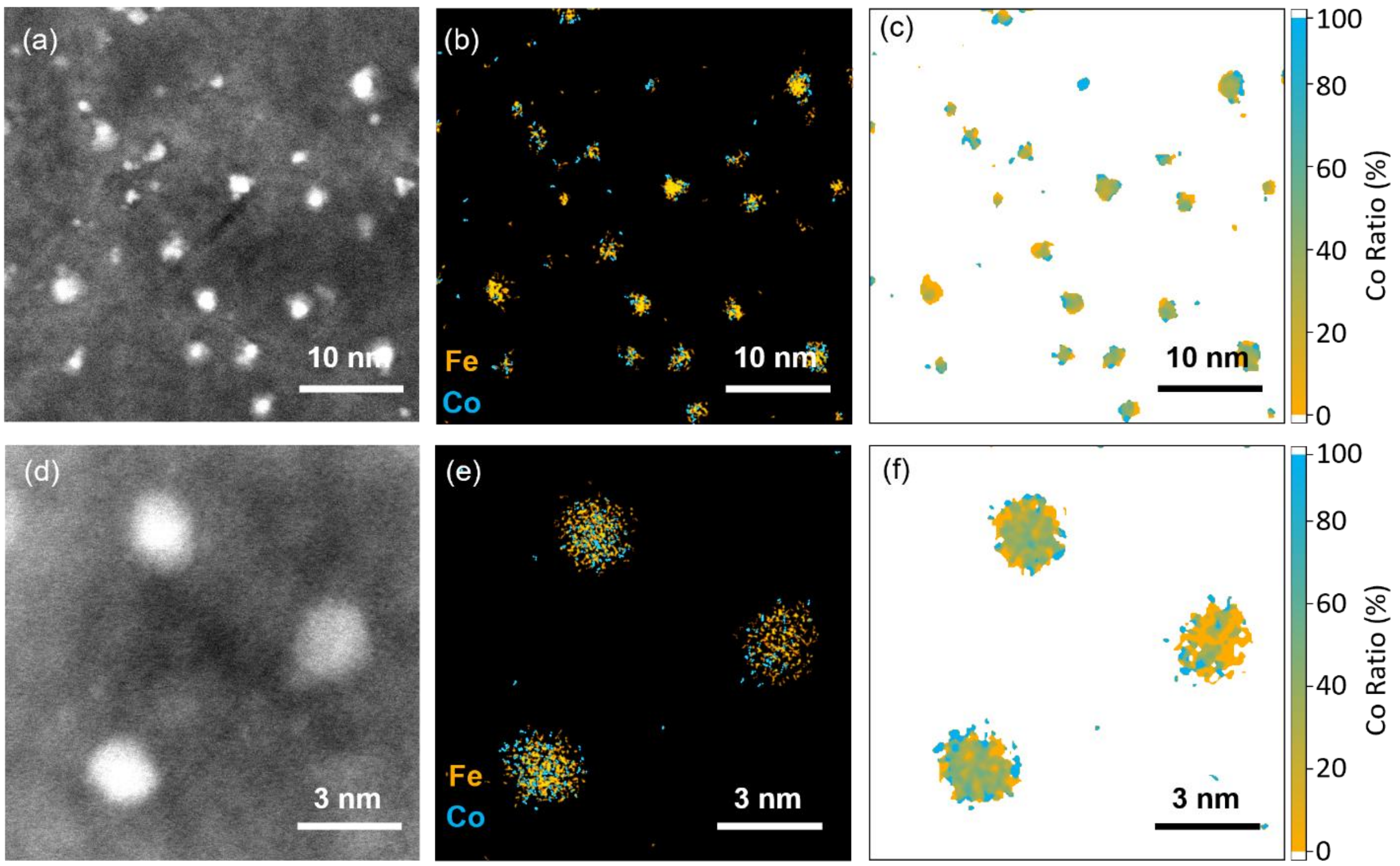

**Figure S12.** STEM characterization and analysis of catalysts containing 25% Co subjected to ACCVD at 850 °C and 50 Pa. (a-c) Low-magnification (large-area) and (d-f) high-magnification (small-area) STEM images. (a) and (d) show annular dark-field (ADF) images; (b) and (e) are the EDS mapping of Fe and Co distributions; (c) and (f) display calculated Co ratio mapping.

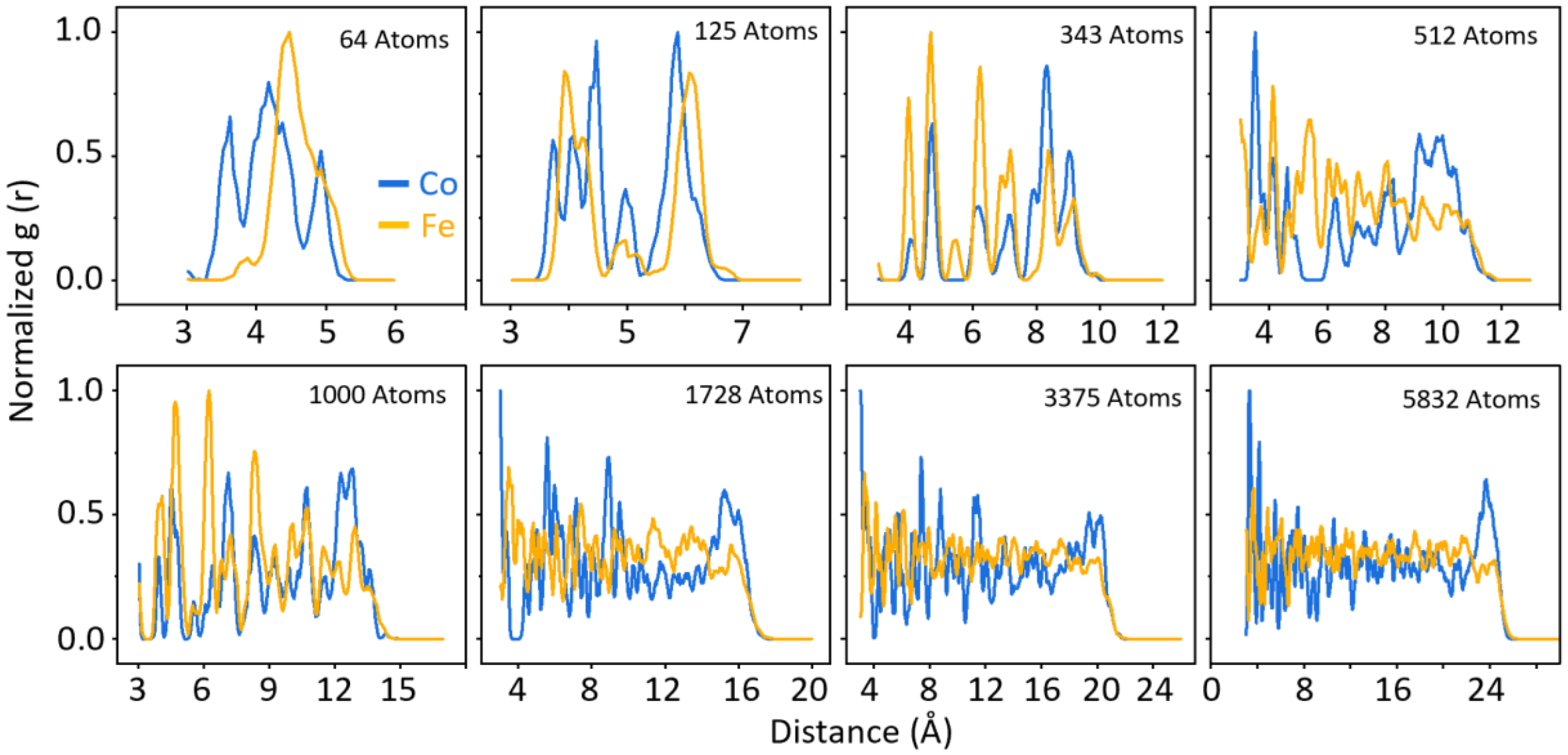

**Figure S13.** Radial distribution profiles of Fe and Co for simulated $Fe_{0.75}Co_{0.25}$ nanoparticles with different total atom numbers, showing size-dependent compositional variations along the radius.

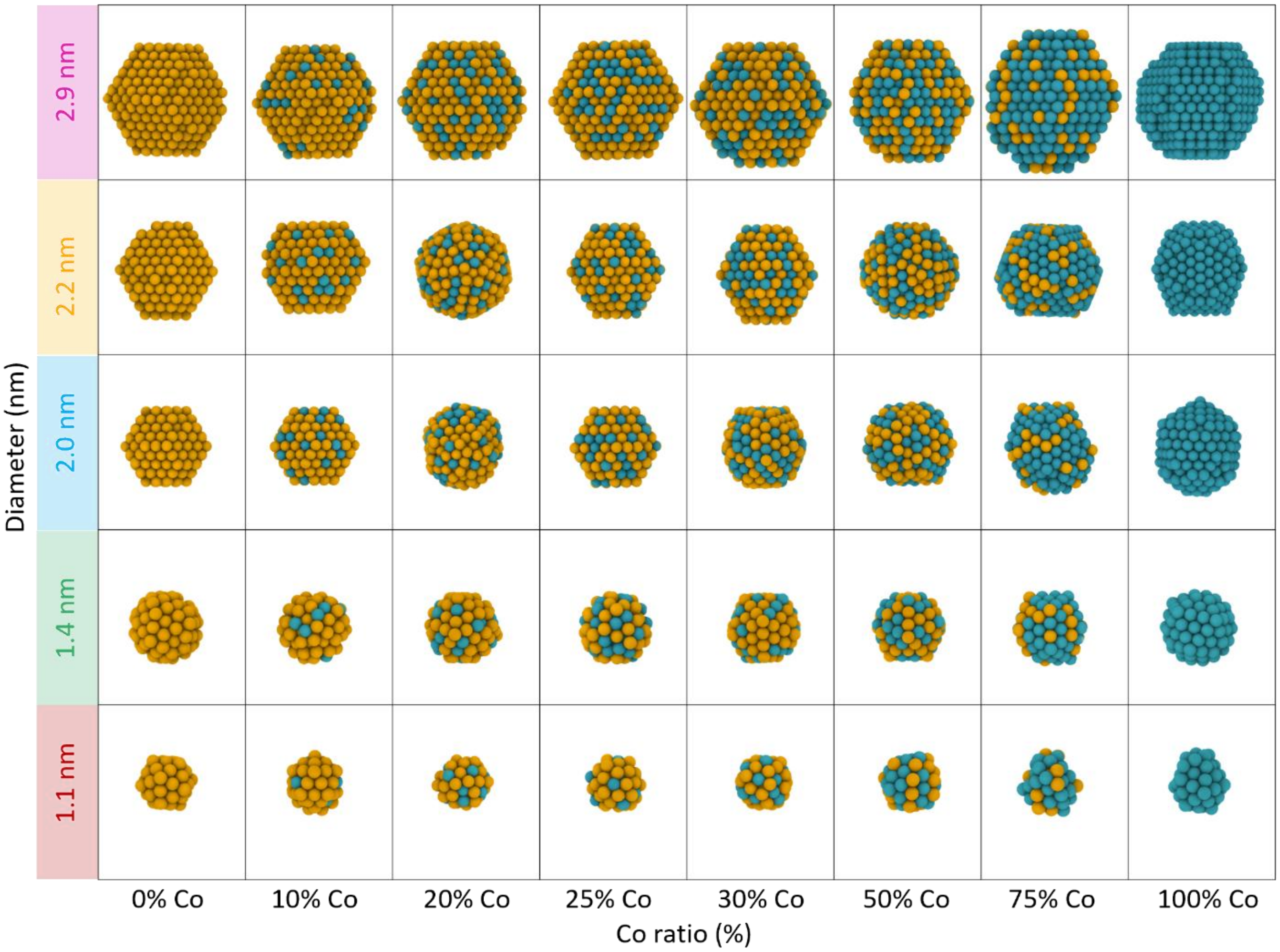

**Figure S14.** Atomic models of Fe-Co clusters: Different compositions of the clusters with 1000 atoms (2.9 nm), 512 atoms (2.2 nm), 343 atoms (2.0 nm), 125 atoms (1.4 nm) and 50 atoms (1.1 nm).

## 2. Detailed fitting process and results of absorption spectra

The entire fitting procedure consists of two main parts: two-dimensional (2D) fitting of fluorescence-excitation (PLE) maps and absorbance (Abs) spectral fitting. The 2D fitting of PLE was performed using the fitting software developed by Cambré *et al.*[3] From this analysis, we systematically extracted the peak positions (emission wavelength ($S_{11}$) and excitation wavelength ($S_{22}$)), full width at half maximum (FWHM), and other relevant parameters for each chirality, which were subsequently converted to the parameters for subsequent absorbance fitting (the detailed conversion procedure is described in **Section 2.3**). The converted parameters were then applied in the peak fitting of the absorbance spectra (that is the experimental Abs. spectrum mentioned in **Fig. 2a** and **2c** in the main text), where Lorentzian functions were used as the fitting model. The detailed processes are described as follows.

### 2.1 PLE fitting

The experimentally obtained 2D PLE spectra were imported into the PLE fitting software. This tool uses a precise empirical model for the excitation line shape and integrates it with an emission line shape model within a 2D fitting framework, based on the model developed in reference.[3] This approach enables accurate fitting of 2D PLE maps for various SWCNT samples, facilitating the direct extraction of line shape features, including peak positions, linewidths, intensities, and other physical quantities, such as phonon sidebands in both excitation and emission spectra. During the fitting process, appropriate parameter settings and meticulous validation ensured the accuracy and reliability of the fitting results.

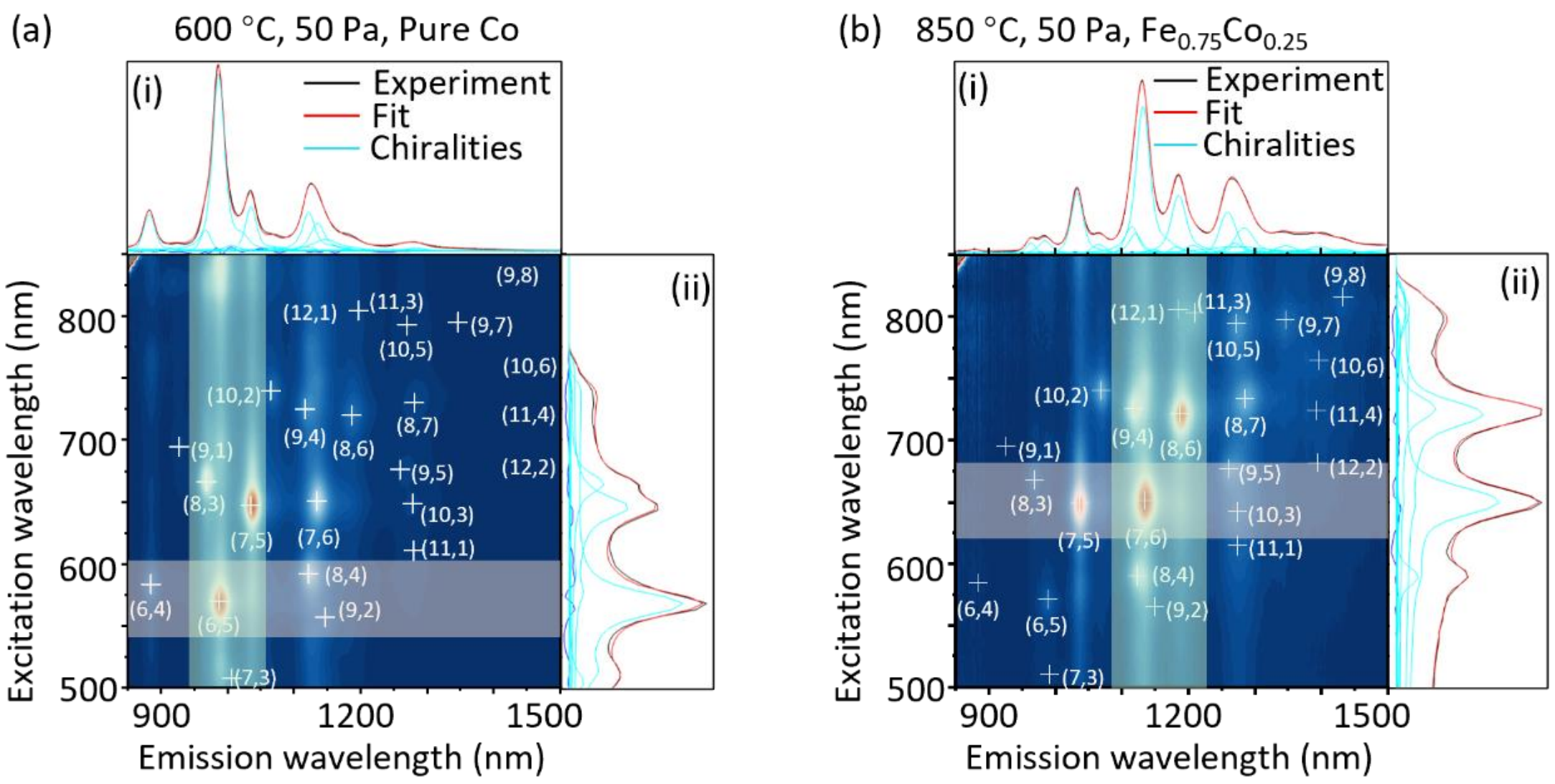

**Figure S15.** The PLE fitting window of SWCNTs grown using (a) pure Co at 600 °C, 50 Pa and (b) $Fe_{0.75}Co_{0.25}$ at 850 °C, 50 Pa. Curve in side-panels show (i) fitted emission spectra (black = experiment, red = total fit and cyan = individual chirality contributions) and (ii) fitted excitation spectra (same color coding).

As shown in **Fig. S15**, the fitting panel provides an intuitive interface that enables users to select specific spectral regions of interest by using a rectangular selection cursor. This functionality facilitates focused and localized evaluation of the fitting results, allowing users to closely inspect individual spectral segments where complex peak overlaps or baseline variations may occur. Furthermore, as illustrated in **Fig. S16**, the overall fitting quality can be comprehensively assessed by comparing the spot size, shape, and precise position of chiral peaks in the fitted spectrum with those in the original experimental data. This comparative analysis ensures that both the intensity distribution and peak resolution are well reproduced by the fitting algorithm.

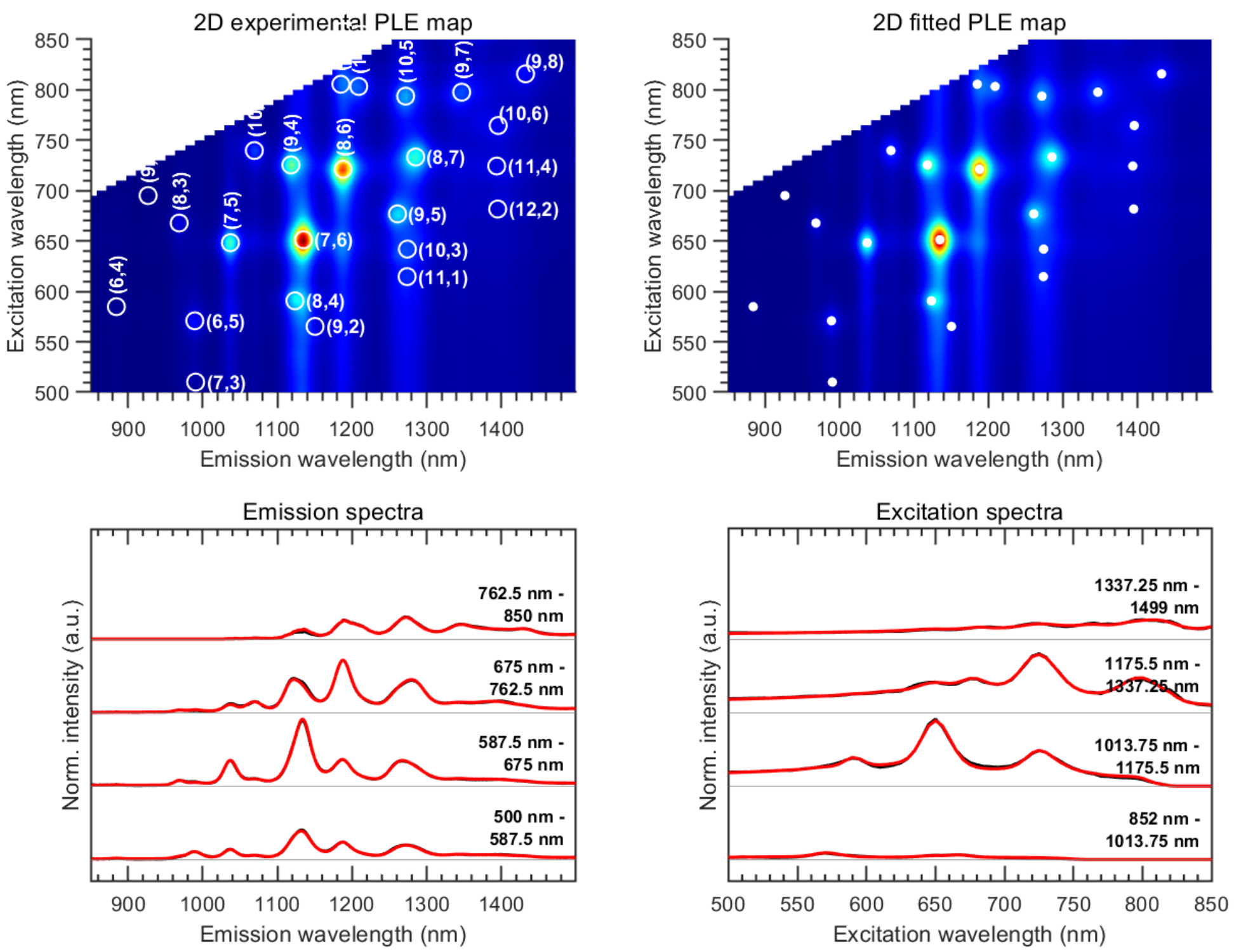

**Figure S16.** Experimental and fitted PLE maps showing excitation and emission spectra by integrating over specific regions of the larger PLE maps, where experimental data (black) and fitted data (red) nicely overlap. The example shown here is for the $Fe_{0.75}Co_{0.25}$ sample grown at 850 °C, 50 Pa. Fitted peaks are indicated by the white markers in the top right panel, while corresponding chiral indices are given in the top left panel.

## 2.2 Absorption spectrum fitting

The peak positions, linewidths, intensities, and phonon sideband parameters obtained from the PLE fitting were converted accordingly and used as initial input parameters for the peak fitting of the absorbance spectra. Lorentzian functions were applied as the fitting model. The final fitting results are presented in **Fig. 2a** and **2c**.

## 2.3 Derivation of the relationship between PLE fitting parameters and those required for absorption fitting

The core of this fitting strategy lies in the rational and effective utilization of the parameters obtained from PLE fitting. To achieve this, a series of systematic verifications were conducted. Initially, the peak positions and linewidths obtained from PLE fitting were directly applied as fixed parameters in the absorbance fitting process. As shown in **Fig. S17a**, comparison of the fitted and original spectra at the (7,5) and (6,5) peaks revealed a noticeable shift of approximately 4 nm towards the blue. This observation indicates that directly using PLE-derived peak positions in the absorbance fitting process leads to systematic deviations, however they can serve as ideal starting parameters for optimizing the absorption fits after applying a general shift, as the relative intensities in the absorption spectrum can already be quite nicely reproduced. Further analysis across multiple samples confirmed the consistency of this shift between absorption and emission peaks, with the degree of shift closely correlated with the type of dispersant used for SWCNT dispersion. The statistical results for different dispersants and organic polymers, including Sodium deoxycholate (DOC), Poly(9,9-dioctylfluorenyl-2,7-diyl)-alt-2,2′-bipyridine (PFO-BPy), and Poly(N-vinylcarbazole) (PCz), are summarized in Table 1. These findings highlight the necessity of shifting the peak positions obtained from PLE fitting before applying them in absorbance fitting. Interestingly, even when applying a single general fixed shift of all the peaks, a noticeable improvement in fitting accuracy can be observed as shown in **Fig. S17b**, though such a fixed shift does not fit all the data perfectly.

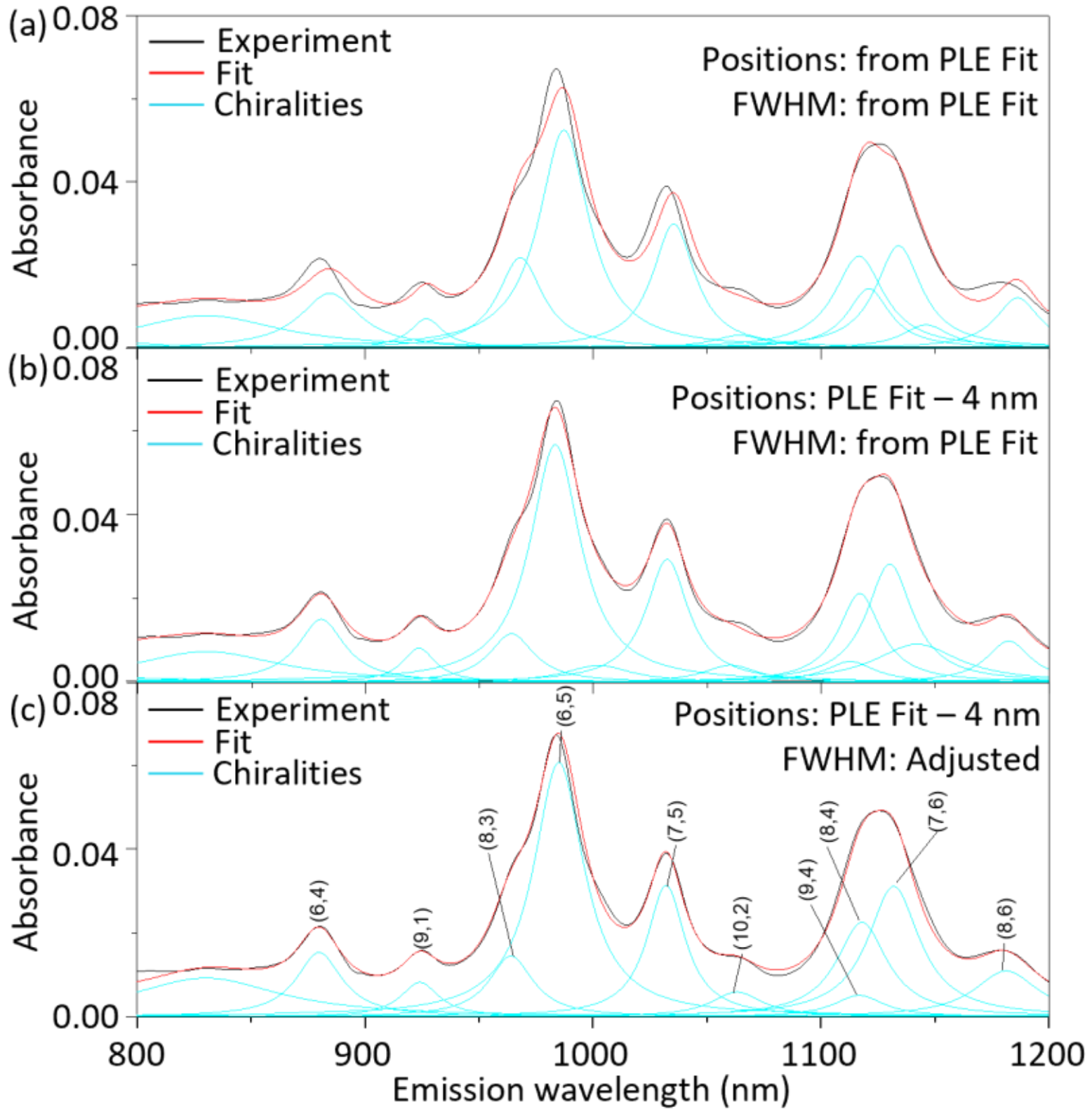

**Figure S17.** Comparison of different fit parameters for the same absorbance spectrum between 800 -1200nm.

Subsequently, we explored the conversion relationship between the linewidth parameters derived from PLE fitting and those required for absorbance fitting. Using the peak positions as fixed parameters, we manually adjusted the linewidths with reference to the PLE fitting results to optimize the absorbance fitting. Note that in the PLE fitting program, a Voigt line shape with corresponding FWHM was used, while here we use Lorentzian line shapes, thus therefore the required adaptation of the line width is logical. The final fitting outcome using the FWHMs as free fit parameters is presented in **Fig. S17c**, showing a very nice correspondence between experiment and fit, without requiring to adjust all peak fitting parameters for each of the individual chiralities which would result in too many fit parameters to realistically optimize. Additionally, the peak strength ratio of $S_{11}$ to $S_{22}$ of each chiral peak obtained by fitting was calculated and plotted as

**Fig. S18.** The peak intensity ratio of $S_{11}$ to $S_{22}$ for most chiralities is found to be clustered around 3.1. This further supports the reliability of our fitting strategy, given it is expected that the ratio should be constant.

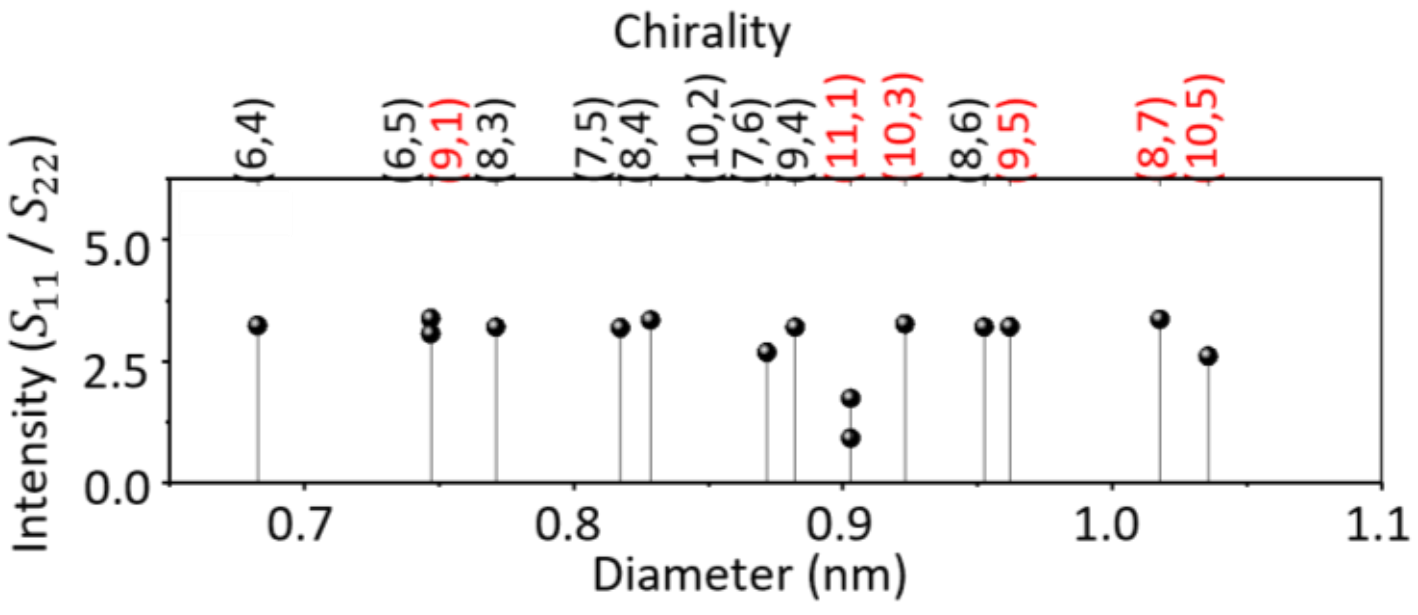

**Figure S18.** Intensity ratio of $S_{11}$ and $S_{22}$ for different chiralities as a function of SWCNT diameter. The chirality with red labels indicates that there might be errors in the assignment.

The extracted linewidth values at this stage were compared with those obtained from PLE fitting, revealing an approximately constant ratio between the two sets of parameters, as illustrated in **Fig. S19**. Moreover, to further validate our approach of using a fixed shift of all peak positions, we also performed an additional test where we fitted the spectra releasing the peak positions as free parameters, allowing them to be optimized for each individual peak, showing that the values did not change much, with a very similar overall fit as a result. Notably, the differences in emission peak positions between the PLE fitting and absorption fitting results remain consistent, with a stable offset of approximately 4 nm. This indicates that applying an appropriate and fixed wavelength correction is essential prior to absorption fitting in the $S_{11}$ region. In contrast, the differences in excitation peak positions between PLE fitting and absorption fitting are essentially negligible, suggesting that no wavelength shift occurs between the two fitting methods in the $S_{22}$ region. As expected, the peak positions in the $S_{22}$ region should coincide between the absorbance spectrum and the PLE map (absorption vs. *excitation*), as opposed to $S_{11}$ (absorption vs. *emission*), where a Stokes shift is expected. This consistency further confirms the reliability and accuracy of the present fitting procedure.

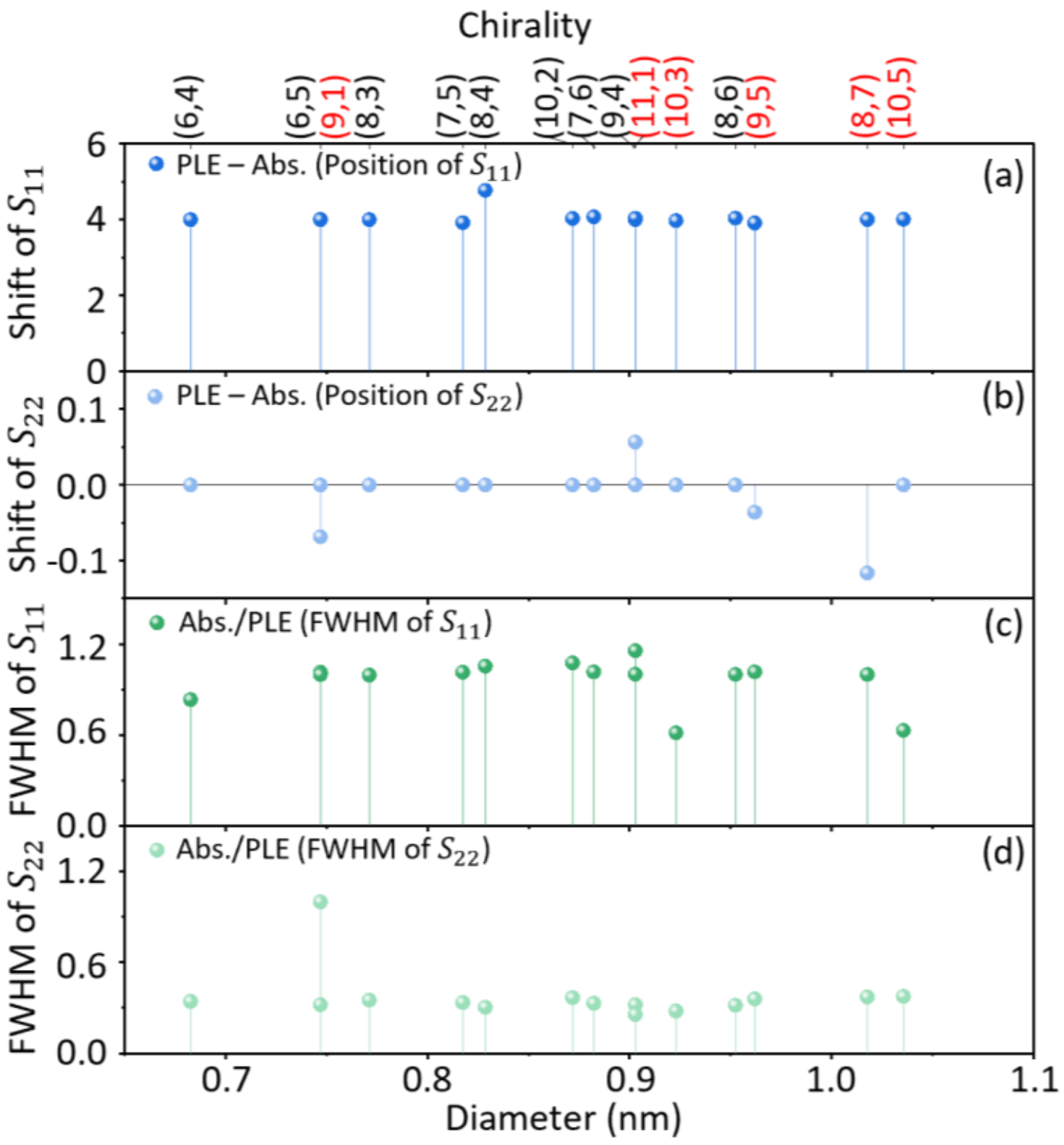

**Figure S19.** The example of preliminary trial fit before we decided to use the constant shift ($S_{11}$ is 4 nm, $S_{22}$ is 0 nm.) extracted parameters of $S_{11}$ and $S_{22}$ for different chiralities, from absorption vs. PLE. Shift of the positions of (a) $S_{11}$and (b) $S_{22}$; The ratio of FWHM of (c) $S_{11}$ and (d) $S_{22}$.

To further assess whether the fitting parameters are influenced by the choice of dispersants used for SWCNT dispersion, we dispersed the same SWCNT sample using different dispersants and performed identical optical characterizations, including absorption spectroscopy and PLE mapping. The fitting procedures were conducted following the same methodology. As shown in **Fig. S20**, the fitting results for the sample dispersed with Poly[(9,9-dioctylfluorenyl-2,7-diyl)-alt-(2,2'-bipyridine-5,5')] (PFO-BPy) are presented. Like the DOC-dispersed system, a consistent emission peak offset between PLE and absorption fitting results was observed. However, the offset for the PFO-BPy system was approximately −8 nm, differing in magnitude and sign from the DOC case. In contrast, the $S_{22}$ excitation peak positions showed negligible differences between PLE and absorption fitting results, again consistent with the trend observed for DOC. These observations collectively demonstrate that the emission peak offset between the two fitting approaches is primarily induced by the dispersant environment, with the offset value depending on the dispersant type. A systematic summary of the emission peak offsets for SWCNTs dispersed in various dispersants is provided in **Table 1** (Readers may refer to this table when performing absorption fitting in subsequent analyses.).

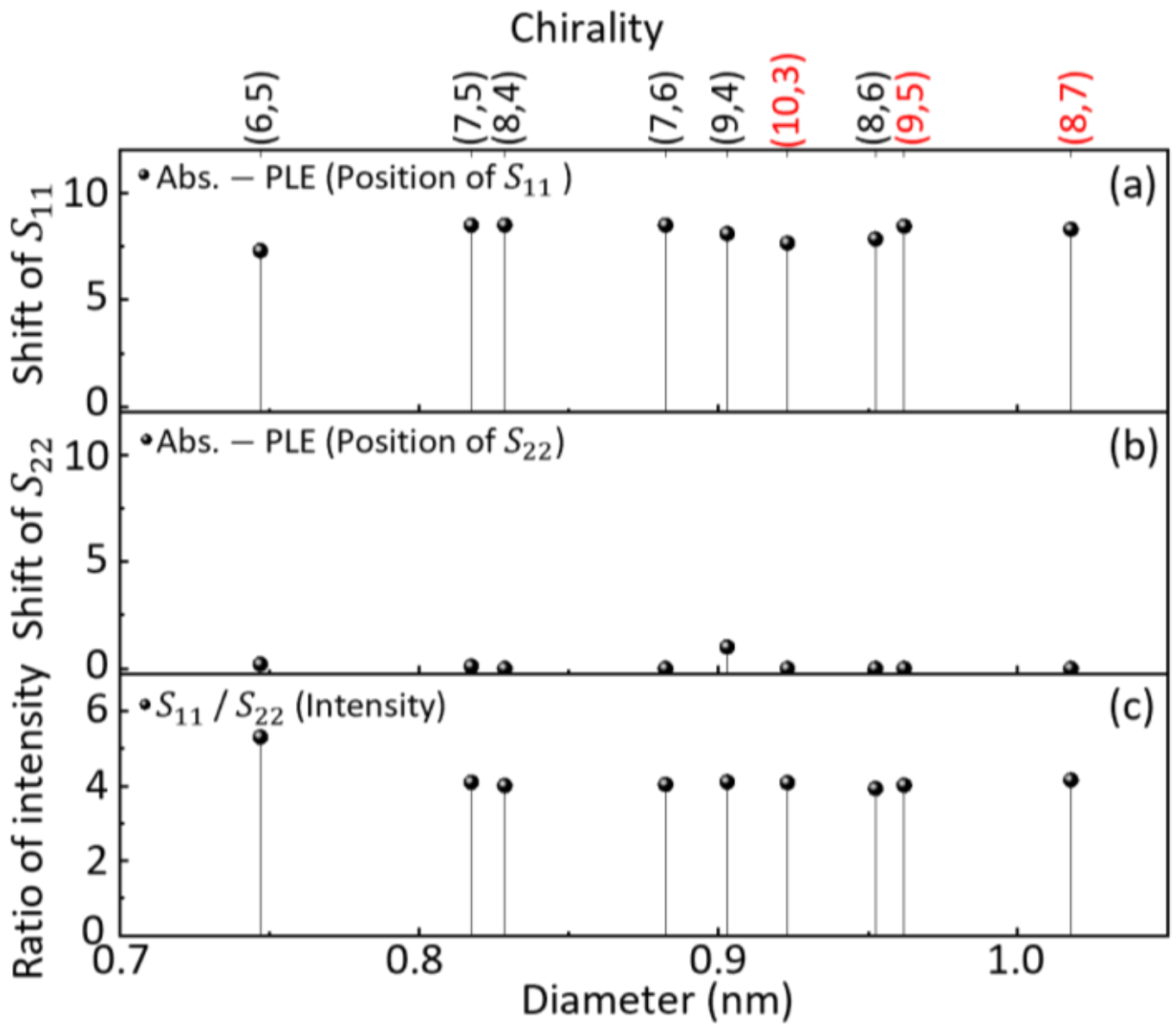

**Figure S20.** The extracted parameters of $S_{11}$ and $S_{22}$ for different chiralities based on PFO-BPy as the dispersant. Shift of the positions of (a) $S_{11}$ and (b) $S_{22}$; (c) The ratio of intensity of $S_{11}$ and $S_{22}$.

**Table S1.** Summary of the recommended offset of $S_{11}$ and $S_{22}$ for the reference of Abs. fitting.

| | DOC | PFO-BPy | PCz |
|---|---|---|---|
| Shift of $S_{11}$ (nm) (PLE – Abs.) | 4 | -8 | -7.4 |
| Shift of $S_{22}$ (nm) (PLE – Abs.) | 0 | 0 | 0 |
| Ratio of FWHM for $S_{11}$ (Abs. / PLE) | 0.95 | 1.29 | 1.23 |
| Ratio of FWHM for $S_{22}$ (Abs. / PLE) | 0.31 | 0.57 | 0.61 |

**2.4. Detailed fitting results of absorbance spectra for pure Co sample grown at 600 °C 50 Pa and $Fe_{0.75}Co_{0.25}$ sample grown at 850 °C 50 Pa**

After establishing the fitting protocol and finalizing all fitting procedures, we proceeded to perform peak deconvolution of the absorption spectrum for the target samples grown under two distinct conditions: 600 °C at 50 Pa and 850 °C at 50 Pa. Based on the fitting results, relative PL efficiencies of different chiral species were calculated using the following equation:

$$Relative\ PL\ Efficiency = \frac{PLE\ Intensity}{Absorption\ Intensity}$$

For this, we use the (7,6) chirality as a normalized reference. Here, we focus on presenting and analyzing the results obtained for the two representative samples discussed in the main text, as shown in **Fig. S21** and **S22**. In the following sections, these two samples are subjected to further comparative analysis and detailed discussion, with particular attention given to the influence of growth temperature on chiral selectivity, peak resolution, and the variation in quantum efficiency across different nanotube species.

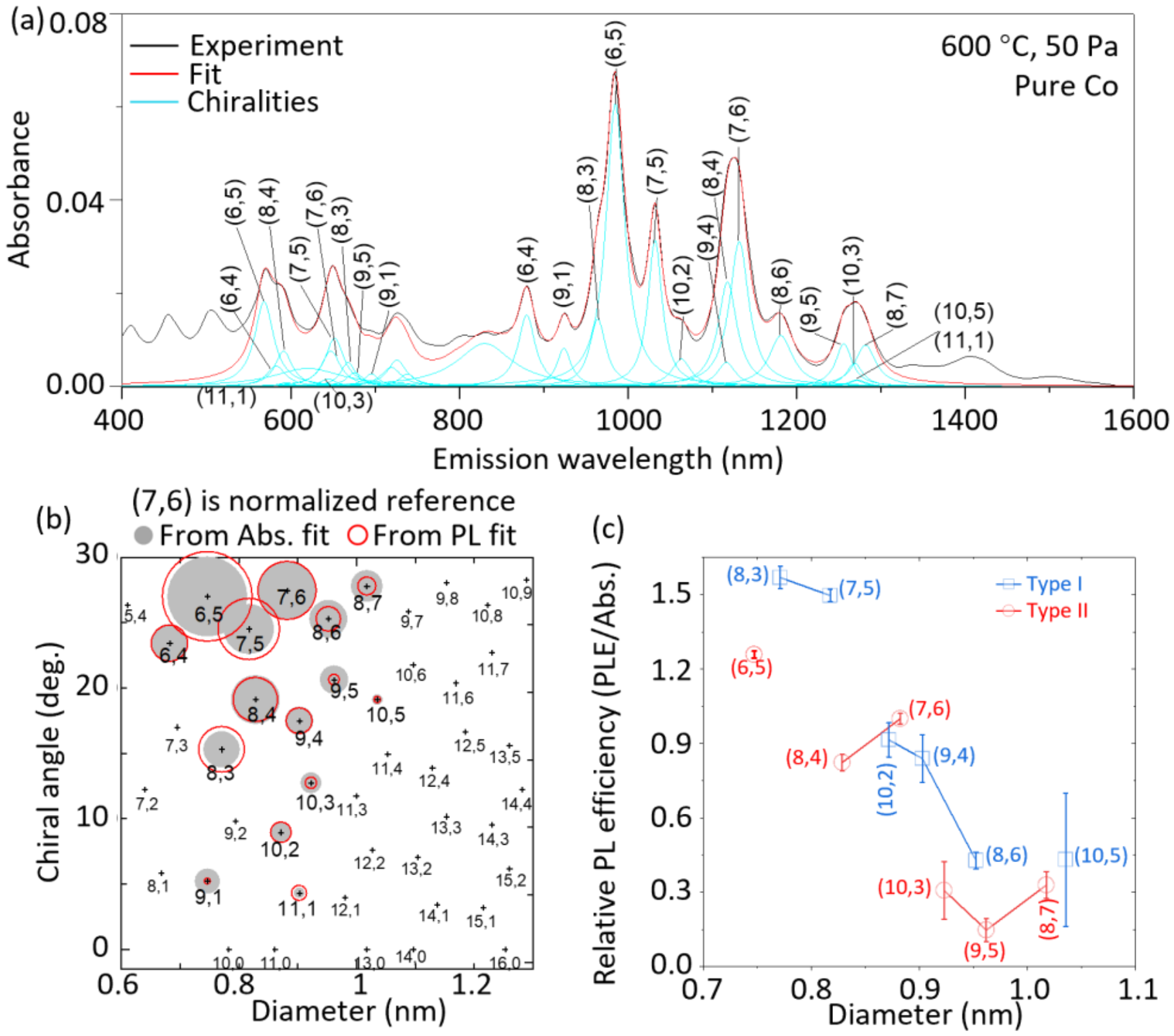

**Figure S21.** (a) Absorption spectrum and corresponding peak fitting results for SWCNTs grown at 600 °C and 50 Pa (pure Co catalyst). The experimental spectrum (black), overall fit (red), and individual chiral contributions (cyan) are shown. (b) 2D chirality distribution map plotted with tube diameter on the x-axis and chiral angle on the y-axis. The circle sizes represent relative intensities obtained from absorption fitting (gray) and PLE fitting (red outlines), with (7,6) used as a normalization of absorption vs. PLE results. (c) Relative PL efficiency (PLE/Abs) as a function of diameter for both Type *I* and Type *II* chiralities, highlighting the higher PL efficiency of small-diameter species. [4-6]

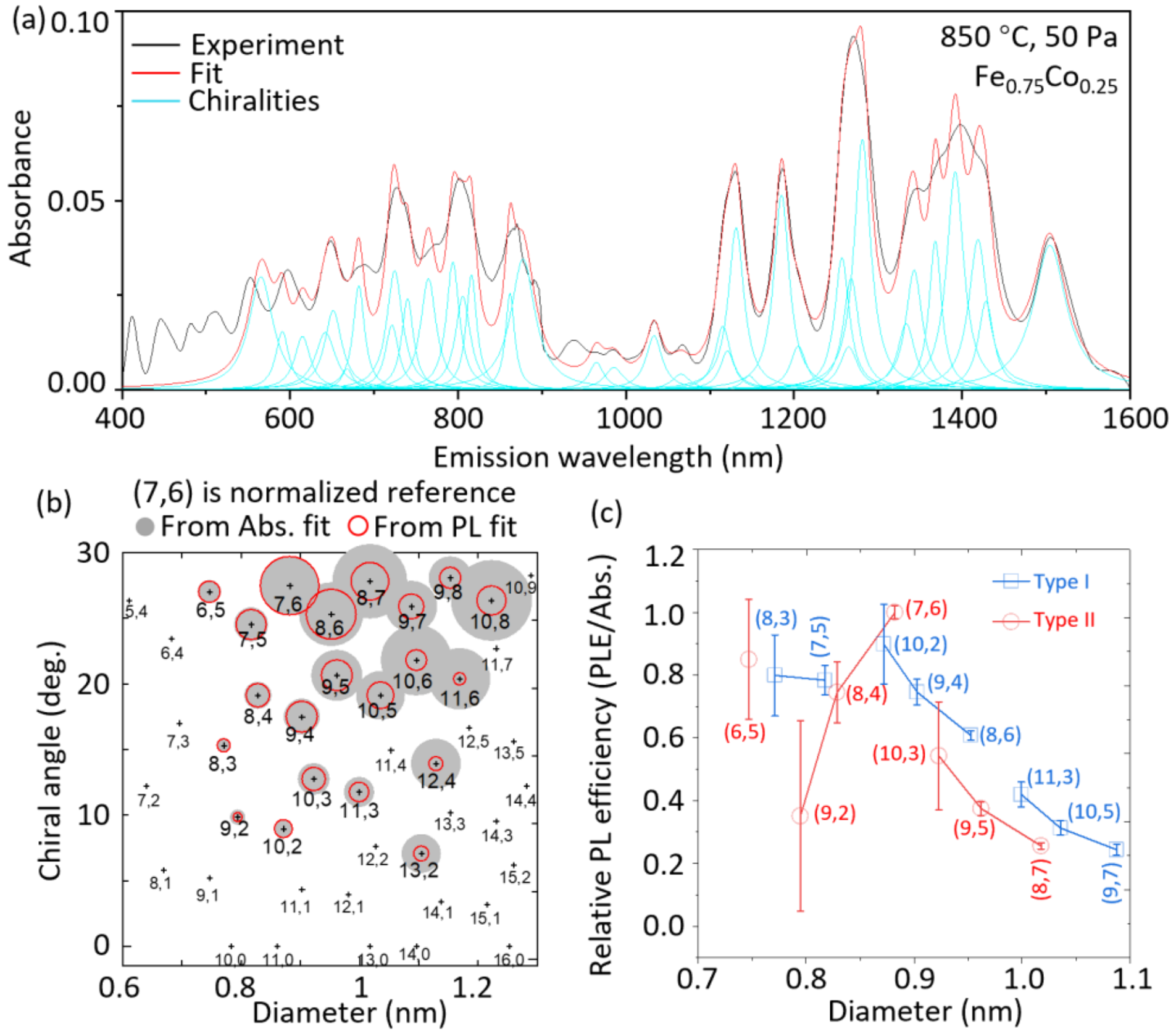

**Figure S22.** (a) Absorption spectrum and corresponding peak fitting results for SWCNTs grown at 850 °C and 50 Pa ($Fe_{0.75}Co_{0.25}$ catalyst). The experimental spectrum (black), overall fit (red), and individual chiral contributions (cyan) are shown. (b) 2D chirality distribution map showing the relative intensities of each chirality, with tube diameter on the x-axis and chiral angle on the y-axis. Circle sizes correspond to relative intensities obtained from absorption fitting (gray) and PLE fitting (red outlines), with (7,6) used as a normalization of absorption vs. PLE results. (c) Relative PL efficiency (PLE/Abs) as a function of diameter, showing reduced PL efficiency and increased fluctuation (error bar) for nanotubes with small diameter compared to those grown at 600 °C. Similarly, there is still a decreased fluctuation (error bar) for nanotubes with large diameter compared to those grown at 600 °C.

Based on the data presented in **Fig. S21** and **S22**, we first verified the reliability of the fitting procedure by comparing the experimental absorption spectra with the fitted results (**Fig. S21a-S22a**). The excellent agreement between the experimental data and the fitted curves demonstrates that the applied fitting deconvolution method accurately captures overlapping peaks and extracts individual chirality intensities. Contributions from metallic SWCNTs (< 600nm) were not included, as they are not present in the PLE maps. Moreover, also a small fraction of larger diameter SWCNTs (>1400 nm in **Fig. S21-S22**) were not included as they were barely visible in the PLE maps and could therefore not be fitted.

Following this validation, we visualized the chirality distribution in a two-dimensional diameter–chiral angle plot (**Fig. S21b** and **S22b**), where the position of each circle corresponds to the geometric parameters of a specific chirality, and the circle size represents the relative intensity derived from absorption (gray) and PLE fitting (red outline). For the sample grown at 600 °C, the most intense chiral species are concentrated in the diameter range of 0.7–0.95 nm and chiral angles of 10–25°, indicating strong chirality selectivity. In contrast, the sample grown at 850 °C exhibits a broader distribution extending to larger diameters and lower chiral angles, reflecting the influence of increased catalyst particle size and altered catalyst activity at higher temperatures. The close match between the distributions derived from PLE and absorption fitting confirms that the fitting method is capable of faithfully reproducing chirality distributions in geometric parameter space.

Finally, the relative quantum yields (defined as the ratio of PLE intensity to absorption intensity) for different chiral species were compared, as shown in **Fig. S21c–S21c** and **S22a**. For the sample grown at 600 °C, small diameter chiralities such as (7,6), (8,4), and (6,5) exhibit relatively high PL efficiencies. Under 850 °C growth conditions, however, the PL efficiencies decrease significantly, and the error bars become noticeably larger across multiple chiralities. Notably, even chiral species that typically display strong PL efficiencies—such as (6,5) and (8,3)—show considerable data scattering. This increased variability can generally be attributed to the amplified relative uncertainty caused by weak absorption peaks and imperfect background subtraction. Interestingly, the family patterns observed in our present results (**Fig. S23a**) complementarily reproduce the calculated PL quantum yield trends reported in a reliable reference (**Fig. S23b**), further confirming the accuracy and robustness of our fitting strategy. A comprehensive summary of peak offsets and PL efficiencies for various dispersants and growth conditions is provided in **Table S1**.

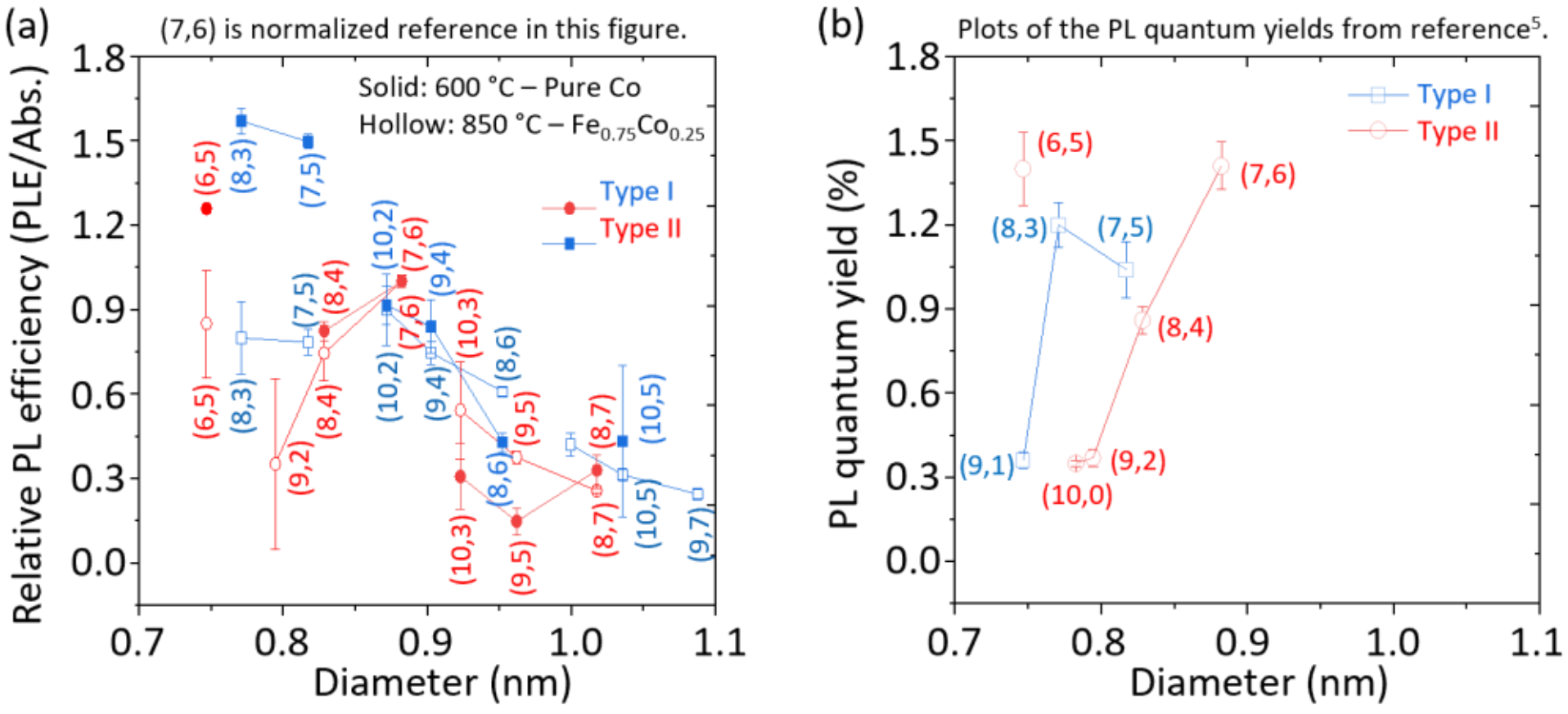

**Figure S23.** Relative PL efficiency (PLE/Abs) as a function of nanotube diameter. (a) Relative PL efficiencies of SWCNTs grown using pure Co at 600 °C (solid symbols) and $Fe_{0.75}Co_{0.25}$ at 850 °C (hollow symbols), combined from **Fig. S21c** and **S22c**. (b) Calculated PL quantum yields of eight (*n*,*m*) SWCNT species as a function of diameter, reproduced from Reference[5].

## REFERENCES

(1) Hou, B.; Wu, C.; Inoue, T.; Chiashi, S.; Xiang, R.; Maruyama, S. Extended Alcohol Catalytic Chemical Vapor Deposition for Efficient Growth of Single-Walled Carbon Nanotubes Thinner than (6,5). *Carbon* **2017**, 119, 502-510.

(2) Tian, Y.; Jiang, H.; Anoshkin, I. V.; Kauppinen, L. J. I.; Mustonenb, K.; Nasibulin, A. G.; Kauppinen, E. I. A Reference Material of Single-walled Carbon Nanotubes: Quantitative Chirality Assessment Using Optical Absorption Spectroscopy. *RSC Advances* **2015**, 5 (125), 102974-102980.

(3) Cambré, S.; Van Werveke, W.; De Clercq, M.; Erkens, M.; Martinati, M.; Wenseleers, W. Quantitative 2D Fitting of Fluorescence-Excitation Maps: Excitation Lineshape of Single-Wall Carbon Nanotubes. *Nanoscale Horizons* **2025**, 10(12), 3405-3415.

(4) Miyauchi, Y.; Chiashi, S.; Murakami, Y.; Hayashida, Y.; Maruyama, S. Fluorescence Spectroscopy of Single-Walled Carbon Nanotubes Synthesized from Alcohol. *Chemical Physics Letters* **2004**, 387 (1-3), 198-203.

(5) Wei, X.; Tanaka, T.; Li, S.; Tsuzuki, M.; Wang, G.; Yao, Z.; Li, L.; Yomogida, Y.; Hirano, A.; Liu, H.; et al. Photoluminescence Quantum Yield of Single-Wall Carbon Nanotubes Corrected for the Photon Reabsorption Effect. *Nano Letters* **2019**, 20 (1), 410-417.

(6) Bachilo, S. M.; Strano, M. S.; Kittrell, C.; Hauge, R. H.; Smalley, R. E.; Weisman, R. B. Structure-Assigned Optical Spectra of Single-Walled Carbon Nanotubes. *Science* **2002**, 298 (5602), 2361-2366.